\pdfoutput=1
\documentclass[12pt,a4paper]{article}

\usepackage{ifthen}
\newboolean{pdflatex}
\setboolean{pdflatex}{true}

\newboolean{articletitles}
\setboolean{articletitles}{true}

\newboolean{uprightparticles}
\setboolean{uprightparticles}{false}

\usepackage{microtype}
\usepackage{lineno}
\usepackage{xspace}
\usepackage{graphicx}
\usepackage{color}
\usepackage{colortbl}
\usepackage{amsmath}
\usepackage{amssymb}
\usepackage{amsfonts}
\usepackage{upgreek}

\newcommand*\patchAmsMathEnvironmentForLineno[1]{%
\expandafter\let\csname old#1\expandafter\endcsname\csname #1\endcsname
\expandafter\let\csname oldend#1\expandafter\endcsname\csname
end#1\endcsname
 \renewenvironment{#1}%
   {\linenomath\csname old#1\endcsname}%
   {\csname oldend#1\endcsname\endlinenomath}%
}
\newcommand*\patchBothAmsMathEnvironmentsForLineno[1]{%
  \patchAmsMathEnvironmentForLineno{#1}%
  \patchAmsMathEnvironmentForLineno{#1*}%
}
\AtBeginDocument{%
\patchBothAmsMathEnvironmentsForLineno{equation}%
\patchBothAmsMathEnvironmentsForLineno{align}%
\patchBothAmsMathEnvironmentsForLineno{flalign}%
\patchBothAmsMathEnvironmentsForLineno{alignat}%
\patchBothAmsMathEnvironmentsForLineno{gather}%
\patchBothAmsMathEnvironmentsForLineno{multline}%
}

\usepackage{hyperref}
\usepackage[all]{hypcap}

\graphicspath{{figures/}}

\def\lhcb {\mbox{LHCb}\xspace}
\def\ux85 {\mbox{UX85}\xspace}

\def\babar  {\mbox{BaBar}\xspace}

\ifthenelse{\boolean{uprightparticles}}%
{

 \def\Pmu         {\ensuremath{\upmu}\xspace}

 \def\Ppi         {\ensuremath{\uppi}\xspace}

 \def\Ppsi        {\ensuremath{\uppsi}\xspace}

 \def\PDelta      {\ensuremath{\Delta}\xspace}                 
 \def\PXi      {\ensuremath{\Xi}\xspace}                 
 \def\PLambda      {\ensuremath{\Lambda}\xspace}                 
 \def\PSigma      {\ensuremath{\Sigma}\xspace}                 
 \def\POmega      {\ensuremath{\Omega}\xspace}                 
 \def\PUpsilon      {\ensuremath{\Upsilon}\xspace}                 
 
 \def\PB      {\ensuremath{\mathrm{B}}\xspace}                 
                  
 \def\PD      {\ensuremath{\mathrm{D}}\xspace}

 \def\PJ      {\ensuremath{\mathrm{J}}\xspace}                 
 \def\PK      {\ensuremath{\mathrm{K}}\xspace}

 \def\Pb      {\ensuremath{\mathrm{b}}\xspace}                 
 \def\Pc      {\ensuremath{\mathrm{c}}\xspace}

 \def\Pi      {\ensuremath{\mathrm{i}}\xspace}

 \def\Ps      {\ensuremath{\mathrm{s}}\xspace}

}
{

 \def\Pmu         {\ensuremath{\mu}\xspace}

 \def\Ppi         {\ensuremath{\pi}\xspace}

 \def\Ppsi        {\ensuremath{\psi}\xspace}                 
                  
 \mathchardef\PDelta="7101
 \mathchardef\PXi="7104
 \mathchardef\PLambda="7103
 \mathchardef\PSigma="7106
 \mathchardef\POmega="710A
 \mathchardef\PUpsilon="7107
                  
 \def\PB      {\ensuremath{B}\xspace}                 
                  
 \def\PD      {\ensuremath{D}\xspace}

 \def\PJ      {\ensuremath{J}\xspace}                 
 \def\PK      {\ensuremath{K}\xspace}

 \def\Pb      {\ensuremath{b}\xspace}                 
 \def\Pc      {\ensuremath{c}\xspace}

 \def\Pi      {\ensuremath{i}\xspace}

 \def\Ps      {\ensuremath{s}\xspace}

}

\def\mumu       {\ensuremath{\Pmu^+\Pmu^-}\xspace}

\def\squark    {\ensuremath{\Ps}\xspace}

\def\cquark    {\ensuremath{\Pc}\xspace}

\def\bquark    {\ensuremath{\Pb}\xspace}

\def\pion  {\ensuremath{\Ppi}\xspace}

\def\pip   {\ensuremath{\pion^+}\xspace}
\def\pim   {\ensuremath{\pion^-}\xspace}
\def\pipi  {\ensuremath{\pion^+\pion^-}\xspace}

\def\kaon  {\ensuremath{\PK}\xspace}
  \def\Kbar  {\kern 0.2em\overline{\kern -0.2em \PK}{}\xspace}

\def\Kz    {\ensuremath{\kaon^0}\xspace}
\def\Kzb   {\ensuremath{\Kbar^0}\xspace}
\def\KzKzb {\ensuremath{\Kz \kern -0.16em \Kzb}\xspace}
\def\Kp    {\ensuremath{\kaon^+}\xspace}
\def\Km    {\ensuremath{\kaon^-}\xspace}

\def\KpKm  {\ensuremath{\Kp \kern -0.16em \Km}\xspace}
\def\KS    {\ensuremath{\kaon^0_{\rm\scriptscriptstyle S}}\xspace}

\def\Kstarzb {\ensuremath{\Kbar^{*0}}\xspace}

  \def\Dbar    {\kern 0.2em\overline{\kern -0.2em \PD}{}\xspace}
\def\D       {\ensuremath{\PD}\xspace}

\def\Dz      {\ensuremath{\D^0}\xspace}
\def\Dzb     {\ensuremath{\Dbar^0}\xspace}
\def\DzDzb   {\ensuremath{\Dz {\kern -0.16em \Dzb}}\xspace}
\def\Dp      {\ensuremath{\D^+}\xspace}
\def\Dm      {\ensuremath{\D^-}\xspace}

\def\DpDm    {\ensuremath{\Dp {\kern -0.16em \Dm}}\xspace}

\def\B       {\ensuremath{\PB}\xspace}
  \def\Bbar    {\kern 0.18em\overline{\kern -0.18em \PB}{}\xspace}

\def\Bz      {\ensuremath{\B^0}\xspace}
\def\Bzb     {\ensuremath{\Bbar^0}\xspace}

\def\Bub     {\ensuremath{\B^-}\xspace}

\def\Bm      {\ensuremath{\Bub}\xspace}

\def\Bd      {\ensuremath{\B^0}\xspace}
\def\Bs      {\ensuremath{\B^0_\squark}\xspace}
\def\Bsb     {\ensuremath{\Bbar^0_\squark}\xspace}
\def\Bdb     {\ensuremath{\Bbar^0}\xspace}

\def\jpsi     {\ensuremath{{\PJ\mskip -3mu/\mskip -2mu\Ppsi\mskip 2mu}}\xspace}

  \def\Y#1S{\ensuremath{\PUpsilon{(#1S)}}\xspace}

\def\Lbar {\ensuremath{\kern 0.1em\overline{\kern -0.1em\PLambda}}\xspace}

\newcommand{\decay}[2]{\ensuremath{#1\!\to #2}\xspace}         

\def\to                 {\ensuremath{\rightarrow}\xspace}

\def\CP                {\ensuremath{C\!P}\xspace}

\def\AT#1     {\ensuremath{A_{\mathrm{T}}^{#1}}\xspace}           

\def\C#1      {\ensuremath{\mathcal{C}_{#1}}\xspace}                       
\def\Cp#1     {\ensuremath{\mathcal{C}_{#1}^{'}}\xspace}                    
\def\Ceff#1   {\ensuremath{\mathcal{C}_{#1}^{\mathrm{(eff)}}}\xspace}        
\def\Cpeff#1  {\ensuremath{\mathcal{C}_{#1}^{'\mathrm{(eff)}}}\xspace}       
\def\Ope#1    {\ensuremath{\mathcal{O}_{#1}}\xspace}                       
\def\Opep#1   {\ensuremath{\mathcal{O}_{#1}^{'}}\xspace}                    

\newcommand{\ket}[1]{\ensuremath{|#1\rangle}}              

\newcommand{\tev}{\ensuremath{\mathrm{\,Te\kern -0.1em V}}\xspace}
\newcommand{\gev}{\ensuremath{\mathrm{\,Ge\kern -0.1em V}}\xspace}
\newcommand{\mev}{\ensuremath{\mathrm{\,Me\kern -0.1em V}}\xspace}
\newcommand{\kev}{\ensuremath{\mathrm{\,ke\kern -0.1em V}}\xspace}
\newcommand{\ev}{\ensuremath{\mathrm{\,e\kern -0.1em V}}\xspace}
\newcommand{\gevc}{\ensuremath{{\mathrm{\,Ge\kern -0.1em V\!/}c}}\xspace}
\newcommand{\mevc}{\ensuremath{{\mathrm{\,Me\kern -0.1em V\!/}c}}\xspace}
\newcommand{\gevcc}{\ensuremath{{\mathrm{\,Ge\kern -0.1em V\!/}c^2}}\xspace}
\newcommand{\gevgevcccc}{\ensuremath{{\mathrm{\,Ge\kern -0.1em V^2\!/}c^4}}\xspace}
\newcommand{\mevcc}{\ensuremath{{\mathrm{\,Me\kern -0.1em V\!/}c^2}}\xspace}

\def\mum  {\ensuremath{\,\upmu\rm m}\xspace}

\def\gsim{{~\raise.15em\hbox{$>$}\kern-.85em
          \lower.35em\hbox{$\sim$}~}\xspace}
\def\lsim{{~\raise.15em\hbox{$<$}\kern-.85em
          \lower.35em\hbox{$\sim$}~}\xspace}

\def\pt         {\mbox{$p_{\rm T}$}\xspace}

\def\pythia     {\mbox{\textsc{Pythia}}\xspace}

\def\gauss      {\mbox{\textsc{Gauss}}\xspace}

\def\tell1  {TELL1\xspace}
\def\ukl1   {UKL1\xspace}

\usepackage{multirow}
\def \t {\theta_{\pi\pi}}
\def \A {{\cal A}}
\usepackage{cite}
\usepackage{mciteplus}

\begin{document}

\renewcommand{\thefootnote}{\fnsymbol{footnote}}
\setcounter{footnote}{1}

\begin{titlepage}
\pagenumbering{roman}

\vspace*{-1.5cm}
\centerline{\large EUROPEAN ORGANIZATION FOR NUCLEAR RESEARCH (CERN)}
\vspace*{1.5cm}
\hspace*{-0.5cm}
\begin{tabular*}{\linewidth}{lc@{\extracolsep{\fill}}r}
\ifthenelse{\boolean{pdflatex}}
{\vspace*{-2.7cm}\mbox{\!\!\!\includegraphics[width=.14\textwidth]{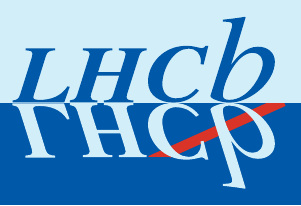}} & &}%
{\vspace*{-1.2cm}\mbox{\!\!\!\includegraphics[width=.12\textwidth]{lhcb-logo.eps}} & &}%
\\
 & & CERN-PH-EP-2013-004 \\
 & & LHCb-PAPER-2012-045 \\
 & & January 24, 2013\\
 \end{tabular*}

\vspace*{3.0cm}

{\bf\boldmath\huge
\begin{center}
Analysis of the resonant components in $\Bdb\rightarrow \jpsi \pi^+\pi^-$
\end{center}
}

\vspace*{1.0cm}

\begin{center}
The LHCb collaboration\footnote{Authors are listed on the following pages.}
\end{center}

\vspace{\fill}

\begin{abstract}
\noindent
Interpretation of \CP violation measurements using charmonium decays,  in both the \Bzb and \Bsb systems, can be subject to changes due to ``penguin" type diagrams. These effects can be investigated using measurements of the Cabibbo-suppressed  $\Bdb\rightarrow \jpsi \pi^+\pi^-$ decays. The final state composition of this channel is investigated
using a 1.0\,{fb}$^{-1}$ sample of data produced in 7 TeV $pp$ collisions at the LHC and collected by the LHCb experiment. A modified Dalitz plot analysis is performed using both the invariant mass spectra and the decay angular distributions.
An improved measurement of the $\Bzb\to\jpsi\pi^+\pi^-$ branching fraction of $(3.97\pm 0.09 \pm 0.11 \pm 0.16)\times 10^{-5}$ is reported where the first uncertainty is statistical, the second is systematic and the third is due to the uncertainty of the branching fraction of the decay $B^- \to \jpsi K^-$ used as a normalization channel. 
Significant production of $f_0(500)$ and $\rho(770)$ resonances is found in the substructure of the $\jpsi \pi^+\pi^-$ final state, and this indicates that they are viable final states for \CP violation studies.
In contrast evidence for the $f_0(980)$ resonance is  not found. This allows us to establish
the first upper limit on the branching fraction product ${\cal{B}}\left(\Bzb\to \jpsi f_0(980)\right)\times{\cal{B}}\left(f_0(980)\to \pi^+\pi^-\right)  < 1.1\times10^{-6}$, leading to an upper limit on the absolute value of the mixing angle of the $f_0(980)$ with the $f_0(500)$ of less than $31^{\circ}$, both at 90\% confidence level.
\end{abstract}

\vspace*{1.0cm}


\begin{center}
Submitted to Physical Review D
\end{center}

\vspace{\fill}

{\footnotesize
\centerline{\copyright~CERN on behalf of the \lhcb collaboration, license \href{http://creativecommons.org/licenses/by/3.0/}{CC-BY-3.0}.}}
\vspace*{2mm}
\end{titlepage}
\newpage
\setcounter{page}{2}

\centerline{\large\bf LHCb collaboration}
\begin{flushleft}
\small
R.~Aaij$^{38}$, 
C.~Abellan~Beteta$^{33,n}$, 
A.~Adametz$^{11}$, 
B.~Adeva$^{34}$, 
M.~Adinolfi$^{43}$, 
C.~Adrover$^{6}$, 
A.~Affolder$^{49}$, 
Z.~Ajaltouni$^{5}$, 
J.~Albrecht$^{9}$, 
F.~Alessio$^{35}$
% ...

{\footnotesize \it
$ ^{1}$Centro Brasileiro de Pesquisas F\'{i}sicas (CBPF), Rio de Janeiro, Brazil\\
% ...
$ ^{5}$Clermont Universit\'{e}, Universit\'{e} Blaise Pascal, CNRS/IN2P3, LPC, Clermont-Ferrand, France\\
$ ^{6}$CPPM, Aix-Marseille Universit\'{e}, CNRS/IN2P3, Marseille, France\\
% ...
$ ^{9}$Fakult\"{a}t Physik, Technische Universit\"{a}t Dortmund, Dortmund, Germany\\
% ...
$ ^{11}$Physikalisches Institut, Ruprecht-Karls-Universit\"{a}t Heidelberg, Heidelberg, Germany\\
% ...
$ ^{33}$Universitat de Barcelona, Barcelona, Spain\\
$ ^{34}$Universidad de Santiago de Compostela, Santiago de Compostela, Spain\\
$ ^{35}$European Organization for Nuclear Research (CERN), Geneva, Switzerland\\
% ...
$ ^{38}$Nikhef National Institute for Subatomic Physics, Amsterdam, The Netherlands\\
% ...
$ ^{43}$H.H. Wills Physics Laboratory, University of Bristol, Bristol, United Kingdom\\
% ...
$ ^{49}$Oliver Lodge Laboratory, University of Liverpool, Liverpool, United Kingdom\\
% ...
$ ^{n}$LIFAELS, La Salle, Universitat Ramon Llull, Barcelona, Spain\\
% ...
}
\end{flushleft}

%
\cleardoublepage
\renewcommand{\thefootnote}{\arabic{footnote}}
\setcounter{footnote}{0}

\pagestyle{plain}
\setcounter{page}{1}
\pagenumbering{arabic}

\section{Introduction}
\label{sec:Introduction}

\begin{figure}[h]
\vskip -.4cm
\begin{center}
\includegraphics[width=5.8in]{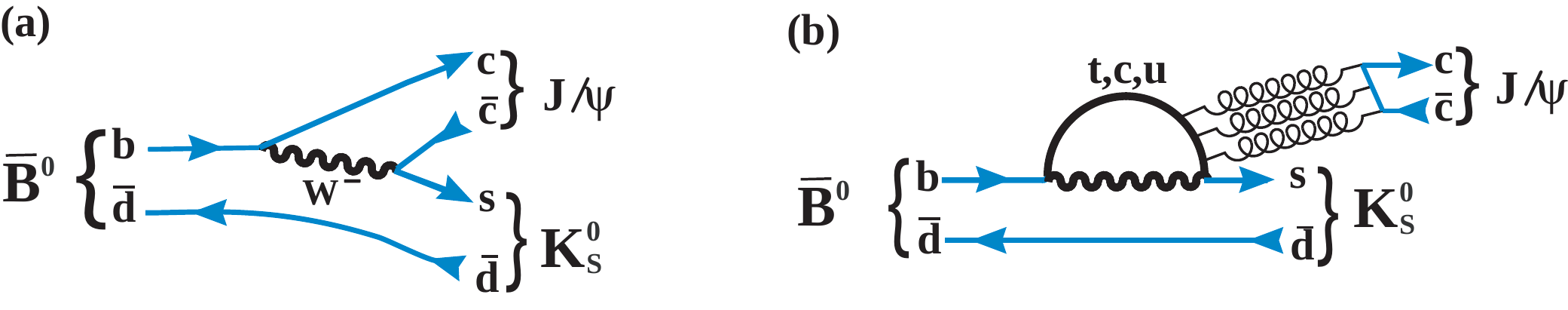}
\end{center}\label{feyn4}
\vskip -0.8cm
\caption{(a) Tree level and (b) penguin diagram examples for $\Bzb$ decays into $\jpsi \KS$.}
\end{figure}

\CP violation measurements using neutral $B$ meson decays into \jpsi mesons are of prime importance both for
determinations of Standard Model (SM) parameters and searching for physics beyond the SM. In the case of \Bzb decays, the final state $\jpsi \KS$ is the most important for measuring $\sin2\beta$~\cite{Aubert:2009aw,*Adachi:2012et,*:2012ke}, while in the case of \Bsb decays, used to measure $\phi_s$, only the final states $\jpsi\phi$ \cite{LHCb-CONF-2012-002,LHCb:2011aa,Aaltonen:2012ie,*Abazov:2011ry,*:2012fu}, and $\jpsi\pi^+\pi^-$ \cite{LHCb:2012ad} have been used so far, where the largest component of the latter is $\jpsi f_0(980)$\cite{LHCb:2012ae,*Stone:2008ak}. The decay rate for these \jpsi modes is dominated by the color-suppressed tree level diagram, an example of which is shown for \Bzb decays in Fig.~\ref{feyn4}(a), while penguin processes, an example of which is shown in Fig.~\ref{feyn4}(b), are expected to be suppressed. Theoretical predictions on the effects of such ``penguin pollution" vary widely for both \Bzb and \Bsb decays \cite{Lenz:2012mb,*Li:2006vq,*Boos:2004xp,*Ciuchini:2005mg,*Bhattacharya:2012ph,*Fleischer:1999nz,*Faller:2008gt}, so it is incumbent upon experimentalists to limit possible changes in the value of the \CP violating angles measured using other decay modes.

\begin{figure}[b]
\vskip -.4cm
\begin{center}
\includegraphics[width=5.8in]{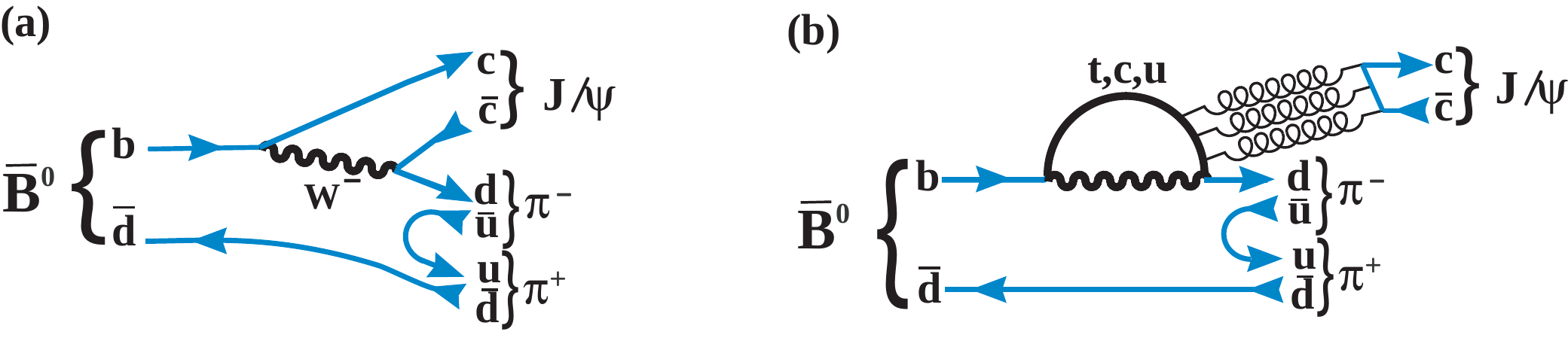}
\end{center}\label{feyn3}
\vskip -.6cm
\caption{(a) Tree level and (b) penguin diagram for $\Bdb$ decays into $\jpsi \pi^+\pi^-$.}
\end{figure}

The decay $\Bzb \to \jpsi \pi^+ \pi^-$ can occur via a Cabibbo suppressed tree level diagram, shown in Fig.~\ref{feyn3}(a), or via several penguin diagrams. An example is shown in Fig.~\ref{feyn3}(b), while others are illustrated in Ref.~\cite{Fleischer:2011au}.  These decays are interesting because they can also be used to measure or limit the amount of penguin pollution. The advantage in using the decay $\Bzb \to \jpsi \pi^+ \pi^-$ arises because the relative amount of pollution is larger. In the allowed decays, e.g. $\Bzb\to \jpsi \KS$, the penguin amplitude is multiplied by a factor of $\lambda^2 R e^{i\phi}$, where $\lambda$ is the sine of the Cabibbo angle $(\approx0.22)$, while in the suppressed decays the factor becomes $R'e^{i\phi'}$, where $R$ and $R'$, and $\phi$ and $\phi'$ are expected to be similar in size \cite{Fleischer:2011au}. A similar study uses the decay $B_s^0\to \jpsi \KS$  \cite{Aaltonen:2011sy,*Aaij:2012di,*DeBruyn:2010hh}.

 \CP violation measurements in the $\jpsi \pi^+ \pi^-$  mode utilizing $\Bz-\Bzb$ mixing determine $\sin2\beta^{\rm eff}$ which can be compared to the well measured $\sin2\beta$. Differences can be used to estimate the magnitude of penguin effects. Knowledge of the final state structure is the first step in this program.
Such measurements on $\sin2\beta^{\rm eff}$ have been attempted in the \Bzb system by using the $\jpsi\pi^0$ final state  \cite{Lee:2007wd,*Aubert:2008bs,*Jung:2012pz}.

In order to ascertain the viability of such \CP violation measurements we perform a full ``Dalitz like" analysis of the final state. 
Regions in $\pi^+\pi^-$ mass that correspond to spin-0 final states would be \CP eigenstates. Final states containing vector resonances, such as the $\rho(770)$ can be analyzed in a similar manner as was done for the decay $\Bsb \to \jpsi \phi$ \cite{LHCb-CONF-2012-002,LHCb:2011aa,Aaltonen:2012ie,*Abazov:2011ry,*:2012fu}.

It is also of interest to search for the $f_0(980)$ contribution and to obtain information concerning the mixing angle between the $f_0(980)$ and the $f_0(500)$\footnote{This particle has been identified previously as the $f_0(600)$ or $\sigma$ resonance.}  partners in the scalar nonet,  as the latter should couple strongly to the $d\bar{d}$ system.
Branching fractions for $\Bzb \to \jpsi \pi^+\pi^-$ and $\jpsi \rho^0$ have previously been measured by the \babar collaboration  \cite{Aubert:2002vb,*Aubert:2007xw}.

In this paper the $\jpsi\pi^+$ and $\pi^+\pi^-$ mass spectra and decay angular distributions are used to determine the resonant and non-resonant components. This differs from a classical Dalitz plot analysis \cite{Dalitz:1953cp} because one of the particles in the final state, the $\jpsi$ meson, has spin-1 and its three decay amplitudes must be considered. We first show that there are no evident structures in the $\jpsi\pi^+$ invariant mass, and then model the $\pi^+\pi^-$ invariant mass with a series of resonant and non-resonant amplitudes. The data are then fitted with the coherent sum of these amplitudes. We report on the resonant structure and the \CP content of the final state.

\section{Data sample and selection requirements}
\label{sec:selections}
The data sample consists of $1.0~\rm fb^{-1}$ of integrated luminosity collected with the \lhcb detector~\cite{LHCb-det} using $pp$ collisions at a center-of-mass energy of 7 TeV. The detector is a single-arm forward
spectrometer covering the pseudorapidity range $2<\eta <5$, designed
for the study of particles containing \bquark or \cquark quarks. Components include a high precision tracking system consisting of a
silicon-strip vertex detector surrounding the $pp$ interaction region,
a large-area silicon-strip detector located upstream of a dipole
magnet with a bending power of about $4{\rm\,Tm}$, and three stations
of silicon-strip detectors and straw drift-tubes placed
downstream.  The combined tracking system has a momentum\footnote{We work in units where $c=1$.}  resolution
$\Delta p/p$ that varies from 0.4\% at 5\gev to 0.6\% at 100\gev,
and an impact parameter resolution of 20\mum for tracks with large
transverse momentum ($\pt$) with respect to the proton beam direction. Charged hadrons are identified using two
ring-imaging Cherenkov (RICH) detectors. Photon, electron and hadron
candidates are identified by a calorimeter system consisting of
scintillating-pad and preshower detectors, an electromagnetic
calorimeter and a hadronic calorimeter. Muons are identified by a
system composed of alternating layers of iron and multiwire
proportional chambers. The trigger consists of a hardware stage, based
on information from the calorimeter and muon systems, followed by a
software stage that applies a full event reconstruction \cite{Aaij:2012me}.

 Events are triggered by a \decay{\jpsi}{\mumu} decay, requiring two identified muons with opposite charge, $\pt(\mu^\pm)$ greater than 500\mev, an invariant mass within 120\mev of the \jpsi mass~\cite{Beringer:2012}, and form a vertex with a fit $\chi^2$ less than 16. After applying these requirements, there is a large \jpsi signal over a small background \cite{Aaij:2011fx}. Only candidates with dimuon invariant mass between $-$48\mev and +43\mev relative to the observed $\jpsi$ mass peak are selected, corresponding a window of  about $\pm3\sigma$. The requirement is asymmetric because of final state electromagnetic radiation. The two muons subsequently are kinematically constrained to the known $\jpsi$ mass.

Other requirements are imposed to isolate $\Bzb$ candidates with high signal yield and minimum background. This is accomplished by combining the $\jpsi\to\mu^+\mu^-$ candidate with a pair of pion candidates of opposite charge, and then testing if all four tracks form a common decay vertex.
Pion candidates are each required to have $\pt$ greater than 250\mev, and the scalar sum of the two transverse momenta, $\pt(\pi^+)+\pt(\pi^-)$, must be larger than 900\mev. The impact parameter (IP) is the distance of closest approach of a track to the primary vertex (PV). To test for inconsistency with production at the PV, 
the IP $\chi^2$ is computed as the difference between the $\chi^2$ of the PV reconstructed with and without the considered track. Each pion must have an IP $\chi^2$ greater than 9. Both pions must also come from a common vertex with an acceptable $\chi^2$ and form a vertex with the \jpsi with a $\chi^2$ per number of degrees of freedom (ndf) less than 10 (here ndf equals five).
Pion and kaon candidates are positively identified using the RICH system. Cherenkov photons are matched to tracks, the emission angles of the photons compared with those expected if the particle is an electron, pion, kaon or proton, and a likelihood is then computed.  The particle identification makes use of the logarithm of the likelihood ratio comparing two particle hypotheses (DLL).  For pion selection we require DLL$(\pi-K)>-10$.

The four-track \Bzb candidate must have a flight distance of more than 1.5~mm, where the average decay length resolution is 0.17~mm. The
angle between the combined momentum vector of the decay products
and the vector formed from the positions of the PV and
the decay vertex (pointing angle) is required to be less than $2.5^{\circ}$.

Events satisfying this preselection are then further filtered using a multivariate analyzer based on a Boosted Decision Tree (BDT) technique~\cite{Breiman}.
The BDT uses six variable that are chosen in a manner that does not
introduce an asymmetry between either the two muons or the two pions. They are the minimum DLL($\mu-\pi$) of the $\mu^+$ and $\mu^-$, the minimum $\pt$ of the $\pi^+$ and $\pi^-$,
the minimum of the IP $\chi^2$ of the $\pi^+$ and $\pi^-$,
the $\Bzb$ vertex $\chi^2$,
the $\Bzb$ pointing angle, and the $\Bzb$ flight distance. There is discrimination power between signal and background in all of these variables, especially the $\Bzb$ vertex $\chi^2$.

\begin{figure}[h]
\vskip -.2cm
\begin{center}
\includegraphics[scale=0.52]{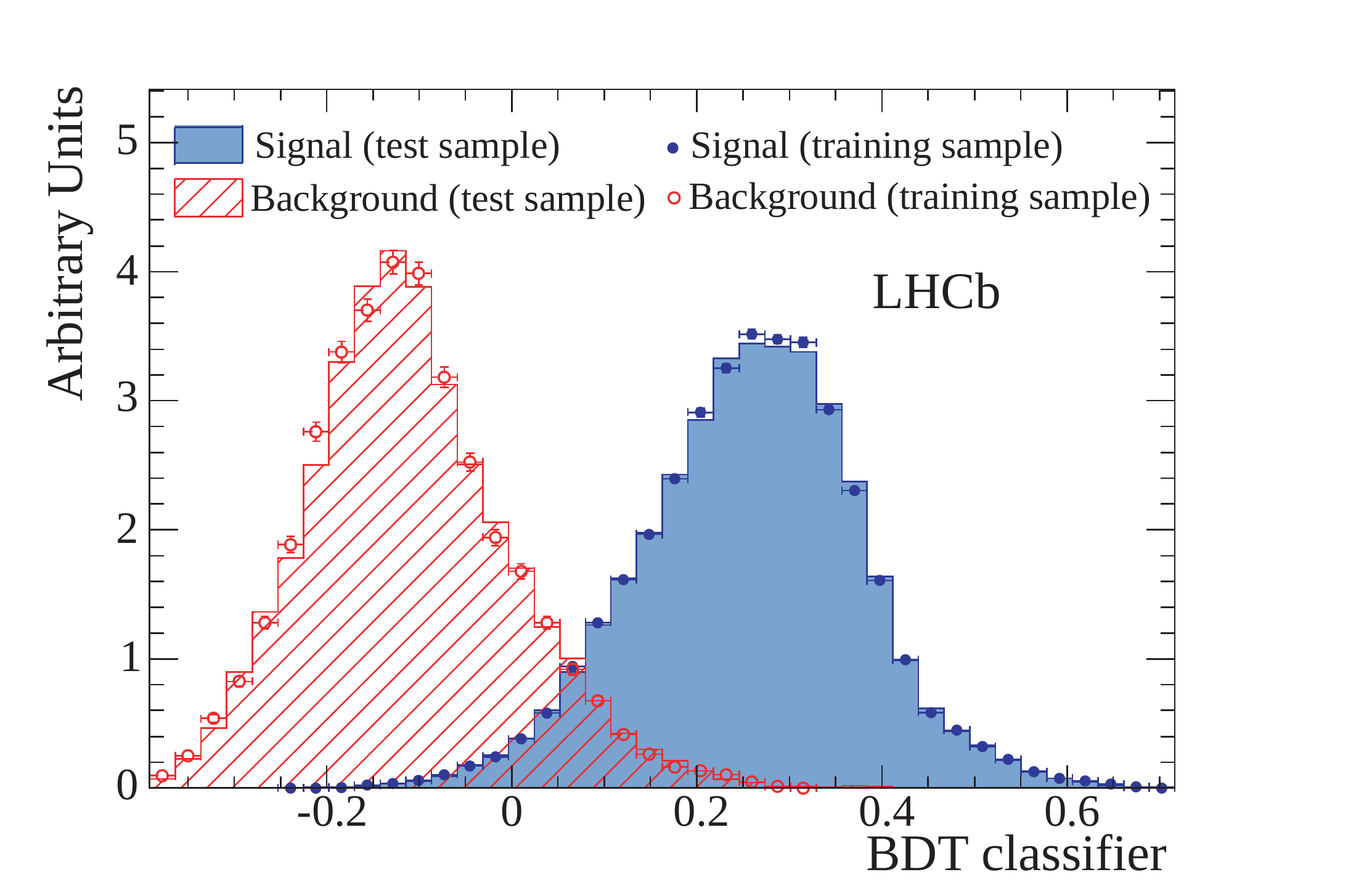}
\end{center}\label{bdt}
\vskip -.8cm
\caption{Distributions of the BDT classifier for both training and test samples of
$\jpsi\pi^+\pi^-$ signal and background events. The signal samples are from simulation and
the background samples are from data.}
\end{figure}

The background sample used to train the BDT consists of the events in the $\Bzb$ mass sideband having $5566 < m(\jpsi \pi^+\pi^-)<5616$\mev. The signal sample consists of two million $\Bdb\rightarrow \jpsi(\rightarrow \mu^+\mu^-)\pi^+\pi^-$ Monte Carlo simulated events that are generated uniformly in phase space, using \pythia \cite{Sjostrand:2006za} with a special LHCb parameter tune \cite{LHCb-PROC-2011-005}, and the LHCb detector simulation based on G{\sc eant}4 \cite{Agostinelli:2002hh} described in Ref.~\cite{LHCb-PROC-2011-006}. Separate samples are used to train and test the BDT. The distributions of the BDT classifier for signal and background are shown in Fig. \ref{bdt}. To minimize a possible bias on the signal acceptance due to the BDT, we choose a relatively loose requirement of the BDT classifier $>0.05$ which has a 96\% signal efficiency and a 92\% background rejection rate.

\begin{figure}[b]
\begin{center}
\vskip -.5cm
\includegraphics[scale=0.55]{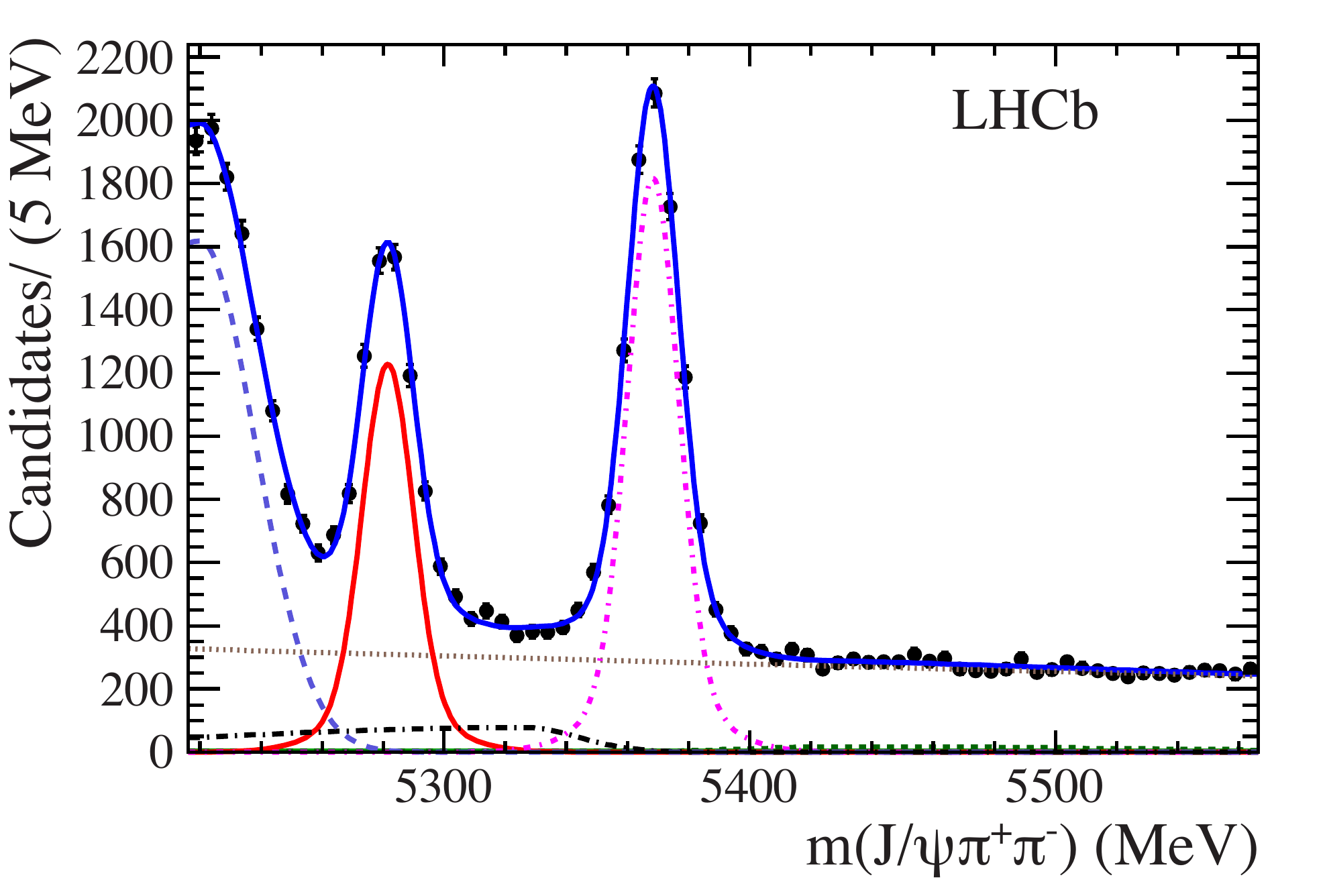}
\end{center}\label{fitmass}
\vskip -0.9cm
\caption{Invariant mass of $\jpsi \pi^+\pi^-$ combinations. The data are fitted with a double-Gaussian signal and several background functions. The (red) solid double-Gaussian function centered at 5280\mev is the $\Bdb$ signal, the (brown) dotted line shows the combinatorial background, the (green) short-dashed shows the $B^-$ background, the (purple) dot-dashed line shows the contribution of $\Bsb\rightarrow \jpsi \pi^+\pi^-$ decays, the (black) dot-long dashed is the sum of $\Bsb\rightarrow \jpsi\eta'(\to \rho \gamma)$ and $\Bsb\rightarrow \jpsi\phi(\to\pi^+\pi^-\pi^0)$ backgrounds, the (light blue) long-dashed is the $\Bdb\rightarrow \jpsi K^- \pi^+$ reflection, and the (blue) solid line is the total.}
\end{figure}

The invariant mass of the selected $\jpsi\pi^+\pi^-$ combinations, where the dimuon pair is constrained to have the $\jpsi$ mass, is shown in Fig.~\ref{fitmass}.  There are signal peaks at both the $\Bsb$ and $\Bzb$ masses on top of the background. 
Double-Gaussian functions are used to fit both signal peaks. They differ only in their mean values, which are determined by the data. The core Gaussian width is also allowed to vary, while the fraction and width ratio of the second Gaussian is fixed to that obtained in the fit of $\Bsb\rightarrow \jpsi \phi$ events. (The details of the fit are given in Ref.~\cite{LHCb:2012ae}.) 
 Other components in the fit model take into account background contributions. One source is from $B^-\rightarrow \jpsi K^-$decays, which contributes when the $K^-$ is misidentifed as a $\pi^-$ and then combined with a random $\pi^+$; the smaller $\jpsi\pi^-$ mode contributes when it is combined with a random $\pi^+$. The next source contains $\Bsb\rightarrow \jpsi\eta'(\rightarrow \rho \gamma)$ and
$\Bsb\rightarrow \jpsi\phi(\rightarrow \pi^+\pi^-\pi^0)$ decays where the $\gamma$ and the $\pi^0$ are ignored respectively. Finally there is a $\Bdb\rightarrow \jpsi K^- \pi^+$ reflection where the $K^-$ is misidentified as $\pi^-$. Here and elsewhere charged conjugated modes are included when appropriate.  The exponential  combinatorial background shape is taken from same-sign combinations, that are the sum of $\jpsi\pi^+\pi^+$ and $\jpsi\pi^-\pi^-$ candidates.
The shapes of the other components are taken from the  simulation with their normalizations allowed to vary. The fit gives $5287\pm112$ signal and $3212\pm80$ background candidates within $\pm20$\mev of the $\Bdb$ mass peak, where a $\KS$ veto, discussed later, is applied.

\begin{figure}[hbt]
\begin{center}
\includegraphics[scale=0.5]{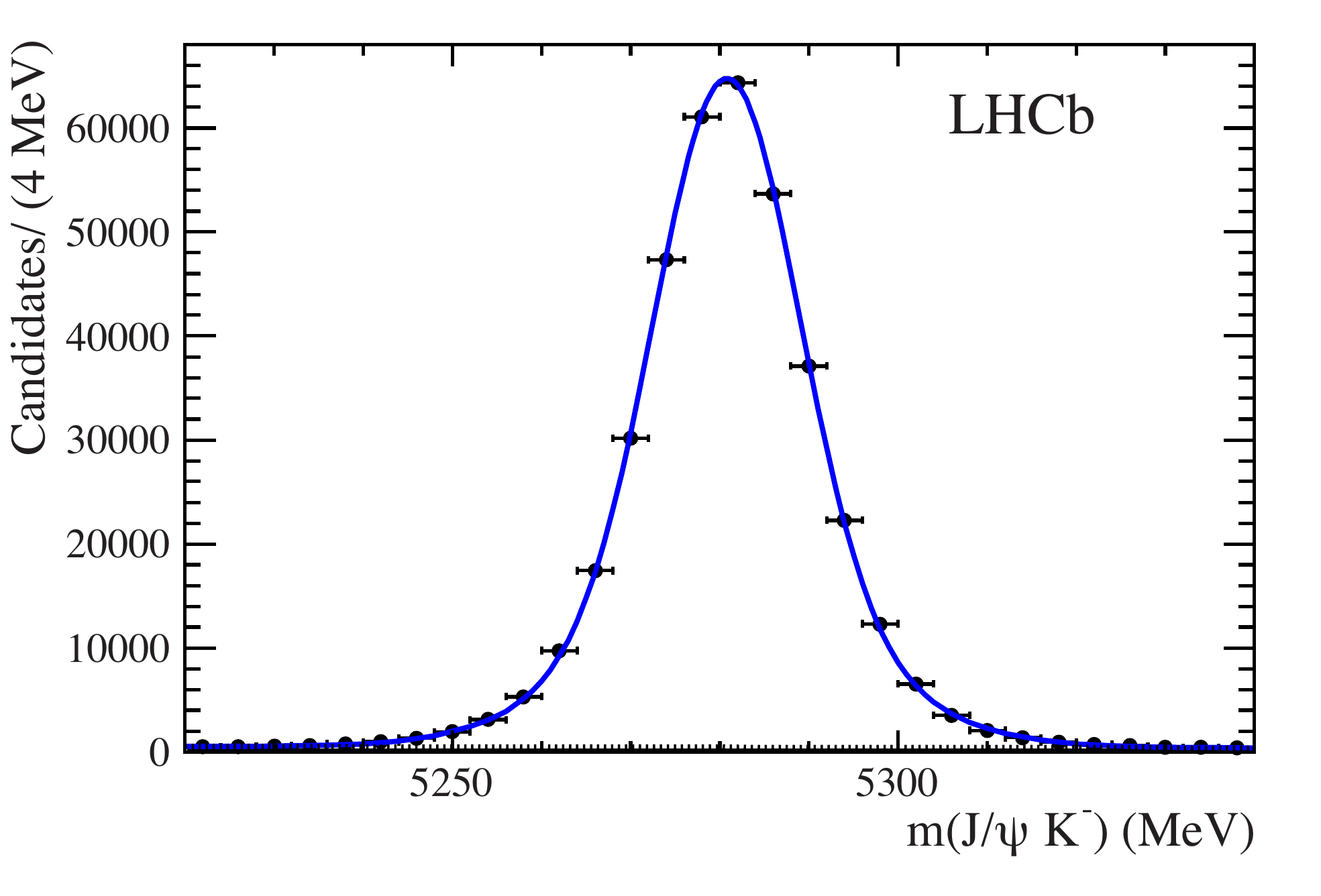}
\end{center}\label{mjpsiK}
\vskip -1cm
\caption{Invariant mass of $\jpsi K^-$ combinations. The data points are fitted with a double-Gaussian function for signal and a linear function for background. The dotted line shows the background, and the (blue) solid line is the total.}
\end{figure}

We use the well measured $B^-\to \jpsi K^-$ mode as a normalization channel to determine the branching fractions.
To minimize the systematic uncertainty from the BDT selection, we employ a similar selection on $B^- \to \jpsi K^-$ decays after requiring the same pre-selection except for particle identification criteria on the $K^-$ candidates. Similar variables are used for the BDT 
except that the variables describing the combination of $\pi^+$ and $\pi^-$ in the $\jpsi \pi^+ \pi^-$ final state are replaced by ones describing the $K^-$ meson. For BDT training, the signal sample uses simulated events and the background sample consists of the data events in the region $5400<m(\jpsi K^-)<5450$\mev. The resulting invariant mass distribution of the candidates satisfying BDT classifier $>0.05$ is shown in Fig. \ref{mjpsiK}. Fitting the distribution with a double-Gaussian function for the signal and linear function for the background gives $350{,}727\pm633$ signal and $4756\pm103$ background candidates within $\pm20$\mev of the $B^-$ mass peak.

\section{Analysis formalism}\label{Formalism}
We apply a formalism similar to that used in Belle's analysis  \cite{Mizuk:2008me} of $\Bzb \to K^-\pi^+\chi_{c1}$ decays and later used in LHCb's analysis of $\Bsb\rightarrow \jpsi \pip \pim$ decays \cite{LHCb:2012ae}.
The decay $\Bdb\rightarrow \jpsi \pi^+\pi^-$, with $\jpsi\rightarrow \mu^+\mu^-$, can be described by four variables. These are taken to be the invariant mass squared of $\jpsi \pi^+$ ($s_{12}\equiv m^2(\jpsi \pi^+)$), the invariant mass squared of $\pi^+\pi^-$ ($s_{23}\equiv m^2(\pi^+\pi^-)$), where we use label 1 for $\jpsi$, 2 for $\pi^+$ and 3 for $\pi^-$, the $\jpsi$ helicity angle ($\theta_{\jpsi}$), which is the angle of the $\mu^+$ in the $\jpsi$ rest frame with respect to the  $\jpsi$ direction in the  $\Bdb$ rest frame,
and the angle between the $\jpsi$ and $\pi^+\pi^-$ decay planes ($\chi$) in the \Bdb rest frame. To improve the resolution of these variables we perform a kinematic fit constraining the $\Bdb$ and $\jpsi$ masses to their nominal values \cite{Beringer:2012}, and recompute the final state momenta.
To simplify the probability density function, we analyze the decay process after integrating over $\chi$, which eliminates several interference terms.

\subsection{\boldmath The decay model for $\Bdb\rightarrow \jpsi \pi^+\pi^-$ }
 The overall probability density function (PDF) given by the sum of signal, $S$, and background, $B$, functions is
\begin{equation}\label{eq:pdf}
F(s_{12}, s_{23}, \theta_{\jpsi})=\frac{f_{\rm sig}}{{\cal{N}}_{\rm sig}}\varepsilon(s_{12}, s_{23}, \theta_{\jpsi}) S(s_{12}, s_{23}, \theta_{\jpsi})+\frac{(1-f_{\rm sig})}{{\cal{N}}_{\rm bkg}} B(s_{12}, s_{23},  \theta_{\jpsi}),
\end{equation}
where  $f_{\rm sig}$ is the fraction of the signal in the fitted region and $\varepsilon$ is the detection efficiency. The fraction of the signal is obtained from the mass fit and is fixed for the subsequent analysis. The normalization factors are given by
\begin{eqnarray}
{\cal{N}}_{\rm sig}&=&\int \! \varepsilon(s_{12}, s_{23}, \theta_{\jpsi}) S(s_{12}, s_{23}, \theta_{\jpsi}) \,
ds_{12}ds_{23}d\cos\theta_{\jpsi},\nonumber\\
{\cal{N}}_{\rm bkg}&=&\int \!B(s_{12}, s_{23}, \theta_{\jpsi}) \,
ds_{12}ds_{23}d\cos\theta_{\jpsi}.
\end{eqnarray}

\begin{figure}[h]
\begin{center}
\includegraphics[width=4.4 in]{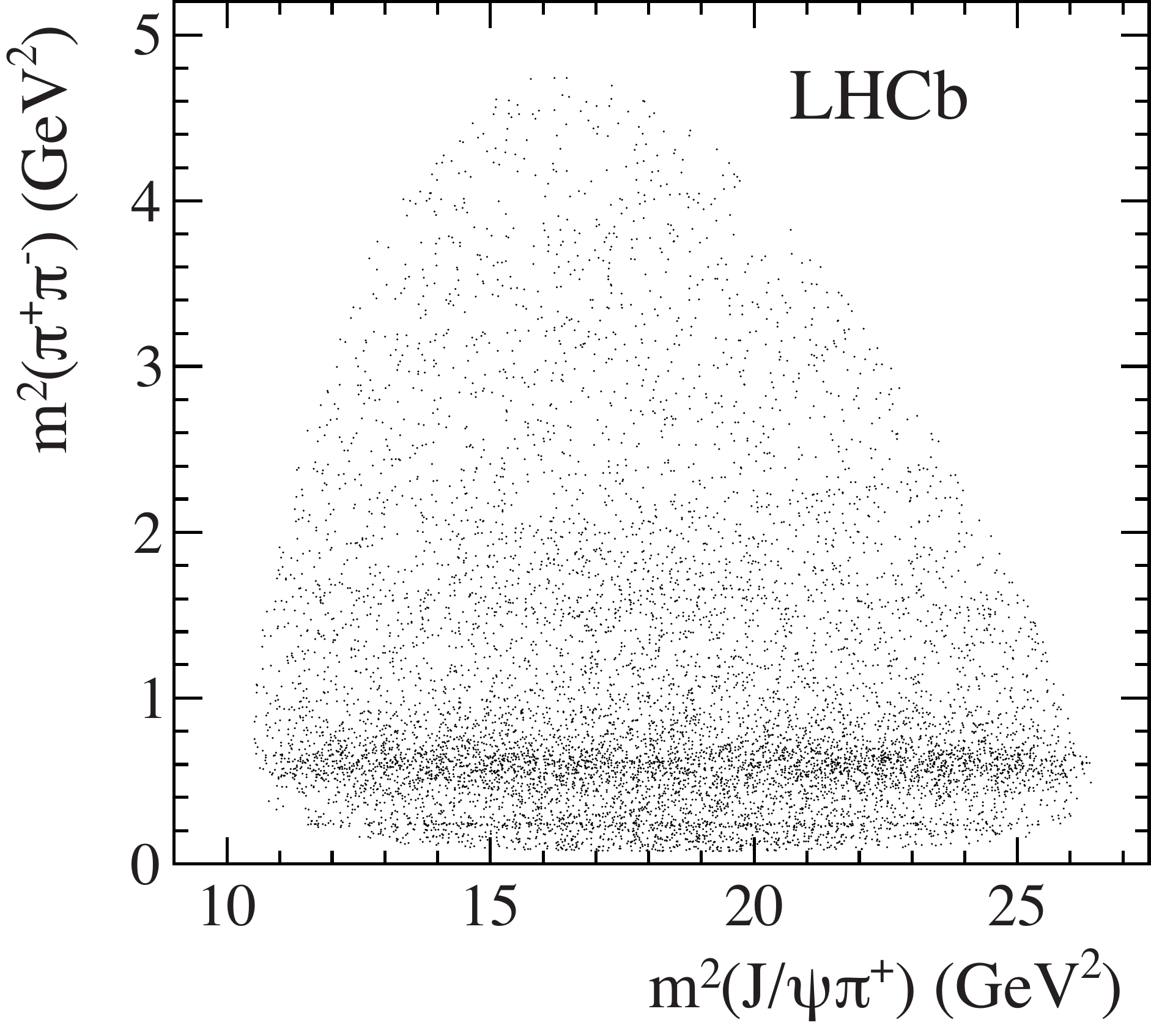}
\caption{Distribution of $m^2(\pi^+\pi^-)$ versus $m^2(\jpsi\pi^+)$ for $\Bdb$ candidate decays within $\pm20$\mev of the  $\Bdb$ mass.}
\end{center}
\label{dalitz-1}
\end{figure}

\begin{figure}[bht]
\begin{center}
\vskip -3mm
\includegraphics[width=0.5\textwidth]{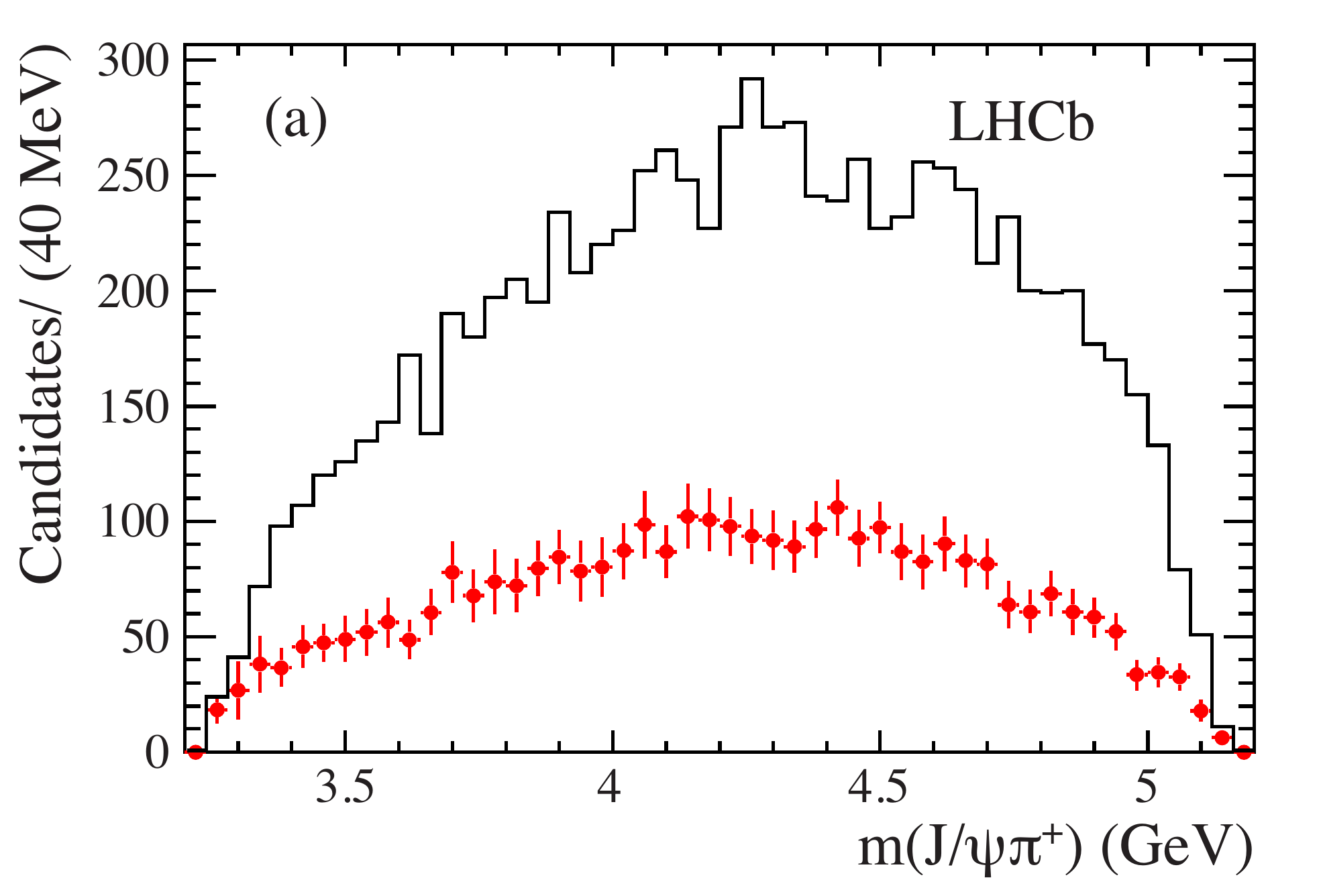}%
\includegraphics[width=0.5\textwidth]{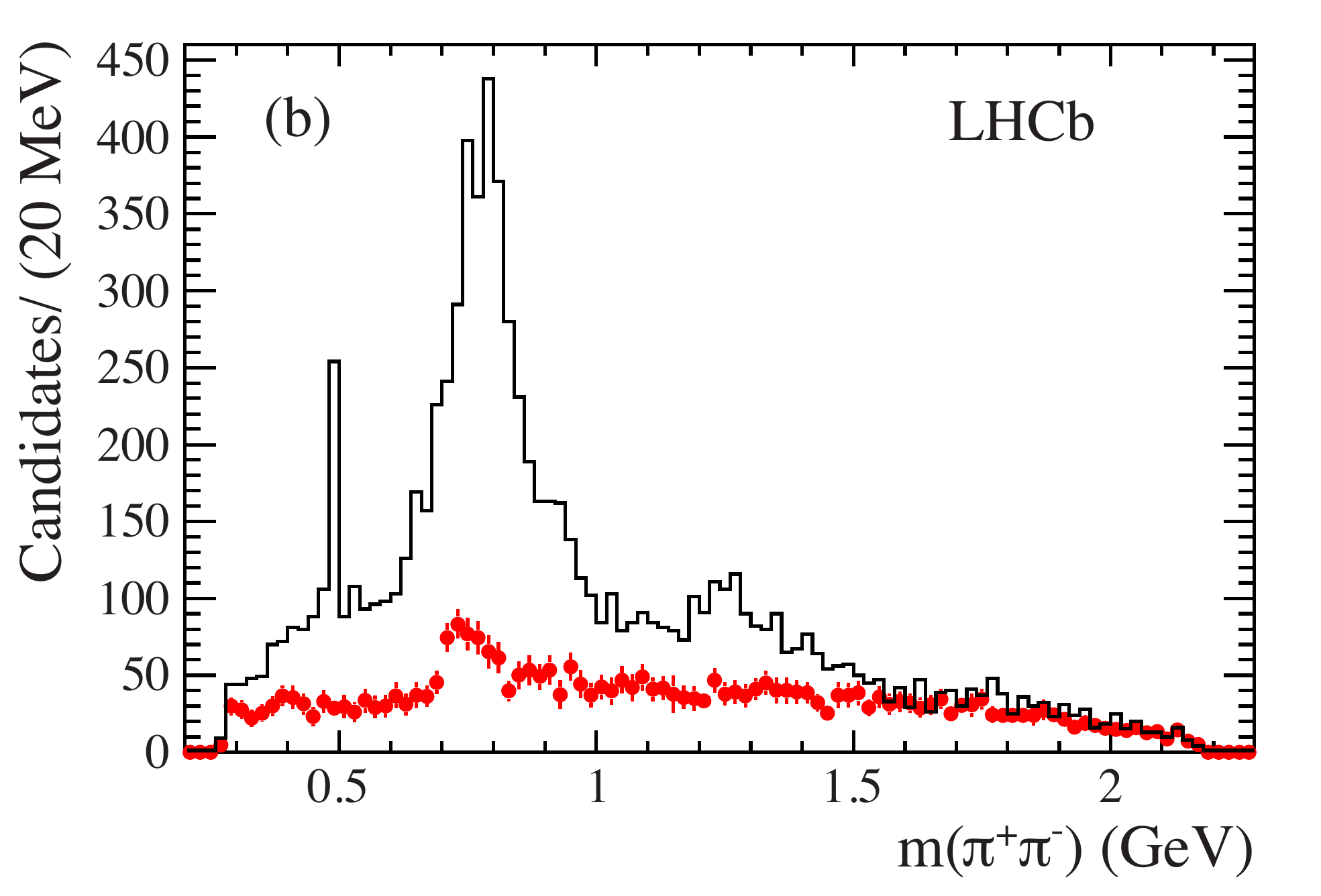}
\caption{Distribution of (a) $m(\jpsi \pi^+)$ and (b) $m(\pi^+\pi^-)$ for $\Bdb\to \jpsi \pi^+\pi^-$ candidate decays within $\pm20$\mev of $\Bdb$ mass shown with the solid line. The (red) points with error bars show the background contribution determined from $m(\jpsi \pi^+\pi^-)$ fits performed in each bin.}
\end{center}
\label{m-jpsipi}
\end{figure}

The event distribution for $m^2(\pi^+\pi^-)$ versus $m^2(\jpsi \pi^+)$ in Fig.~\ref{dalitz-1} shows
obvious structure in $m^2(\pi^+\pi^-)$. To investigate if there are visible exotic structures in the $\jpsi\pi^+$ system as claimed in similar decays \cite{Z4430},
 we examine the $\jpsi \pi^+$  mass distribution shown in Fig.~\ref{m-jpsipi} (a). No resonant effects are evident.
 Figure~\ref{m-jpsipi} (b) shows the $\pi^+\pi^-$ mass distribution. There is a clear peak at the $\rho(770)$ region, a small bump around 1250\mev, but no evidence for the $f_0(980)$ resonance. The favored $\Bdb \to \jpsi \KS$ decay is mostly rejected by the $\Bdb$ vertex $\chi^2$ selection, but about 150 such events remain. We eliminate them by excluding the candidates that have $|m(\pi^+\pi^-)-m_{\KS}|<$~25\mev, where $m_{\KS}$ is the $\KS$ mass \cite{Beringer:2012}.

\subsubsection{The signal function}

The signal function for $\Bzb$ is taken to be the coherent sum over resonant states that can decay into  $\pi^+\pi^-$, plus a possible non-resonant S-wave contribution{\footnote{The interference terms between different helicities are zero because we integrate over the angular variable $\chi$.}
\begin{equation}
S(s_{12}, s_{23}, \theta_{\jpsi})=\sum_{\lambda=0,\pm1}\left|\sum_{i}a^{R_i}_{\lambda}e^{i\phi^{R_i}_{\lambda}}
\mathcal{A}_{\lambda}^{R_i}(s_{12}, s_{23}, \theta_{\jpsi})\right|^2, \label{amplitude-eq}
\end{equation}
where $\mathcal{A}_{\lambda}^{R_i}(s_{12}, s_{23}, \theta_{\jpsi})$ is the amplitude of the decay via an intermediate resonance $R_i$ with helicity $\lambda$. Each $R_i$ has an associated amplitude strength $a_{\lambda}^{R_i}$ for each helicity state $\lambda$ and a phase $\phi_{\lambda}^{R_i}$. Note that the spin-0 component can only have a $\lambda=0$ term. The amplitudes for each $i$ are defined as
\begin{equation}
\mathcal{A}_{\lambda}^R(s_{12},s_{23}, \theta_{\jpsi})= F_B^{(L_B)}\; F_R^{(L_R)}\; A_R(s_{23})\;
 \Big(\frac{P_B}{m_B}\Big )^{L_B}\; \Big( \frac{P_R}{\sqrt{s_{23}}}\Big )^{L_R}\; T_{\lambda}\;\Theta_{\lambda}(\theta_{\jpsi}),\label{eq4}
\end{equation}
where $P_B$ is the \jpsi momentum in the $\Bdb$ rest frame and $P_R$ is the momentum of
either of the two pions in the dipion rest frame, $m_{B}$ is the $\Bdb$ mass, $F_B^{(L_B)}$ and $F_R^{(L_R)}$ are the $\Bdb$ meson and $R$ resonance Blatt-Weisskopf barrier factors~\cite{Blatt}, $L_B$ is the orbital angular momentum between the $\jpsi$ and $\pi^+\pi^-$ system, and $L_R$ is the orbital angular momentum in the $\pi^+\pi^-$ decay and is equal to the spin of resonance $R$ because pions have spin-0. Since the parent $\Bdb$ has spin-0 and the $\jpsi$ is a vector, when the $\pi^+\pi^-$ system forms a spin-0 resonance, $L_B=1$ and $L_R=0$. For $\pi^+\pi^-$ resonances with non-zero spin, $L_B$ can be 0, 1 or 2 (1, 2 or 3) for $L_R=1(2)$ and so on. We take the lowest $L_B$ as the default and consider the other possibilities in the systematic uncertainty.

The Blatt-Weisskopf barrier factors $F_B^{(L_B)}$ and  $F_R^{(L_R)}$ are
 \begin{eqnarray}
F^{(0)} &=& 1, \nonumber \\
F^{(1)} &=& \frac{\sqrt{1+z_0}}{\sqrt{1+z}},\\
F^{(2)} &=& \frac{\sqrt{z_0^2+3z_0+9}}{\sqrt{z^2+3z+9}}. \nonumber
\end{eqnarray}
For the $B$ meson $z = r^2P_B^2$, where the hadron scale $r$ is taken as $5.0\gev^{-1}$, and for the $R$ resonance $z = r^2P_R^2$ with $r$ taken as $1.5\gev^{-1}$ \cite{Kopp:2000gv}. In both cases $z_0= r^2P_0^2$ where $P_0$ is the decay daughter momentum calculated at the resonance pole mass.

The angular term, $T_{\lambda}$, is obtained using the helicity formalism and is defined as
\begin{equation}
 T_{\lambda} = d^J_{\lambda 0}(\theta_{\pi\pi}),
\end{equation}
where $d$ is the Wigner $d$-function,
 $J$ is the resonance spin, $\theta_{\pi\pi}$ is the $\pi^+\pi^-$ resonance helicity angle which is defined as the angle of the $\pi^+$ in the $\pi^+\pi^-$ rest frame with respect to the $\pi^+\pi^-$direction in the $\Bdb$ rest frame  and calculated from the other variables as
 \begin{equation}
\cos \theta_{\pi\pi} = \frac{\left[m^2(\jpsi \pi^+)-m^2(\jpsi \pi^-)\right]m(\pi^+\pi^-)}{4P_R P_B m_{B}}. \label{heli1}
\end{equation}
The $\jpsi$ helicity dependent term  $\Theta_{\lambda}(\theta_{\jpsi})$ is defined as
\begin{eqnarray}
 \Theta_{\lambda}(\theta_{\jpsi})&=& \sqrt{\sin^2\theta_{\jpsi}}\;\;\;\;\;\;\;\;\; \text{for}\;\; \text{helicity} = 0 \nonumber \\
 &=&\sqrt{\frac{1+\cos^2\theta_{\jpsi}}{2}}\;\; \text{for}\;\; \text{helicity} = \pm1. \label{heli2}
\end{eqnarray}

The function $A_R(s_{23})$ describes the mass squared shape of the resonance $R$, that in most cases is a
Breit-Wigner (BW) amplitude. Complications arise, however,  when a new decay channel opens close to the resonant mass. The proximity of a second threshold distorts the line shape of the amplitude.  This happens for the $f_0(980)$  resonance because the $K^+K^-$ decay channel opens. Here we use a Flatt\'e model \cite{Flatte:1976xv} which is described below.

The BW amplitude for a resonance decaying into two spin-0 particles, labeled as 2 and 3, is
\begin{equation}
A_R(s_{23})=\frac{1}{m^2_R-s_{23}-im_R\Gamma(s_{23})}~,
\end{equation}
where $m_R$ is the resonance pole mass, $\Gamma(s_{23})$ is its energy-dependent width that is parametrized as
\begin{equation}
\Gamma(s_{23})=\Gamma_0\left(\frac{P_R}{P_0}\right)^{2L_R+1}\left(\frac{m_R}{\sqrt{s_{23}}}\right)F^2_R~.
\end{equation}
Here $\Gamma_0$ is the decay width when the invariant mass of the daughter combinations is equal to $m_R$.


The Flatt\'e model is parametrized as
\begin{equation}
A_R(s_{23})=\frac{1}{m_R^2-s_{23}-im_R(g_{\pi\pi}\rho_{\pi\pi}+g_{KK}\rho_{KK})}.
\end{equation}
The constants  $g_{\pi\pi}$ and $g_{KK}$ are the $f_0(980)$ couplings to $\pi\pi$ and $K\Kbar$ final states respectively.
 The $\rho$ factors account for the Lorentz-invariant phase space and are given as
\begin{eqnarray}
\rho_{\pi\pi} &=& \frac{2}{3}\sqrt{1-\frac{4m^2_{\pi^{\pm}}}{m^2(\pi^+\pi^-)}}+\frac{1}{3}\sqrt{1-\frac{4m^2_{\pi^{0}}}{m^2(\pi^+\pi^-)}}\label{flatte1}, \\
\rho_{KK} &=& \frac{1}{2}\sqrt{1-\frac{4m^2_{K^{\pm}}}{m^2(\pi^+\pi^-)}}+\frac{1}{2}\sqrt{1-\frac{4m^2_{K^{0}}}{m^2(\pi^+\pi^-)}}.\label{flatte2}
\end{eqnarray}


For non-resonant processes, the amplitude $\mathcal{A}(s_{12},s_{23}, \theta_{\jpsi})$ is derived from Eq.~\ref{eq4}, considering that the $\pi^+\pi^-$ system is S-wave (i.e. $L_R=0$, $L_B=1$) and $A_R(s_{23})$ is constant over the phase space $s_{12}$ and $s_{23}$. Thus, it is parametrized as
\begin{equation}
\mathcal{A}(s_{12},s_{23}, \theta_{\jpsi}) = \frac{P_B}{m_B} \sqrt{\sin^2\theta_{\jpsi}}.
\end{equation}

\subsubsection{Detection efficiency}
\label{sec:mc}

\begin{figure}[b]
\centering
\includegraphics[width =0.48\textwidth]{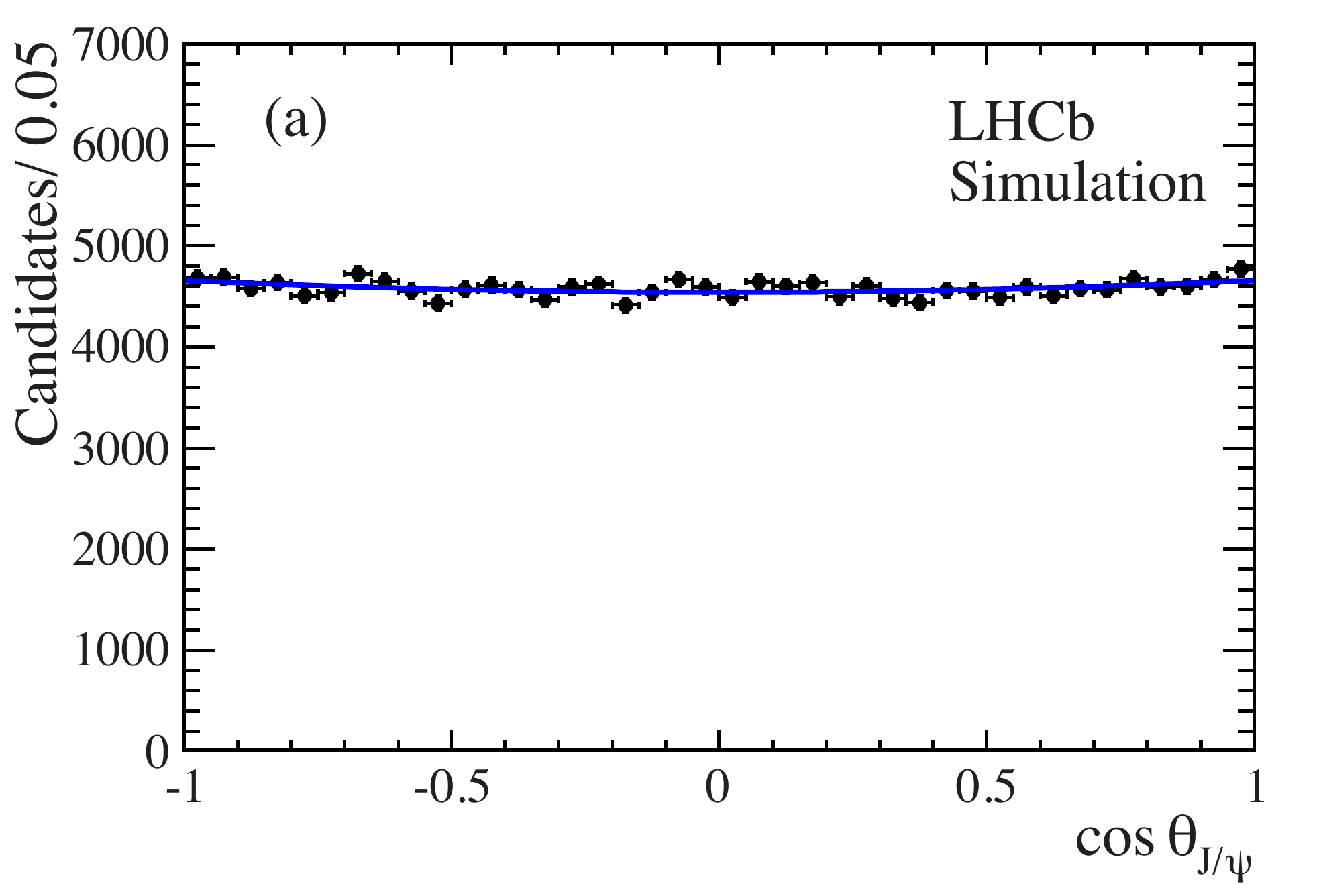}
\includegraphics[width =0.48\textwidth]{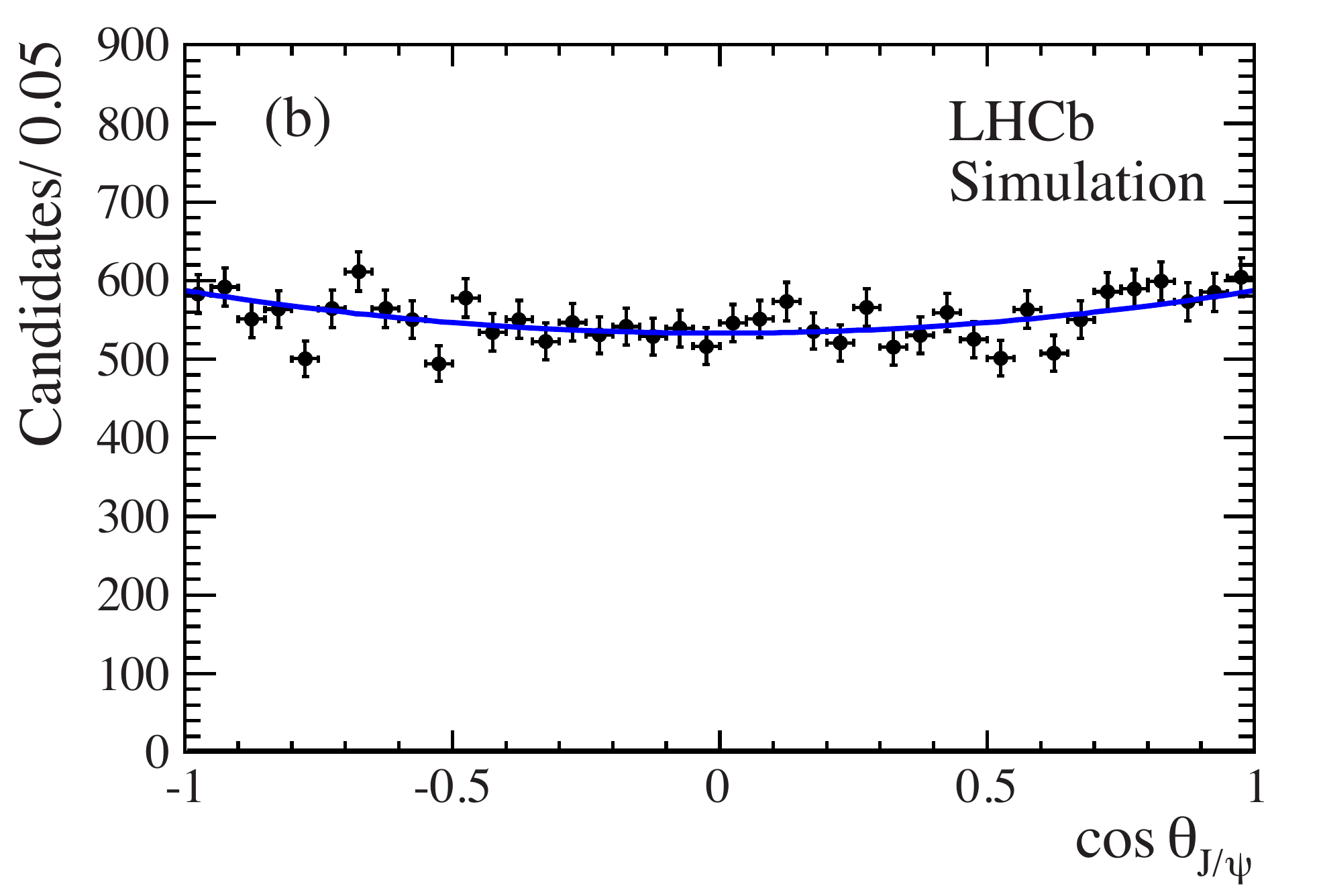}
\caption{Distributions of $\cos\theta_{\jpsi}$ for the $\jpsi \pi^+\pi^-$ simulated sample in (a) the entire dipion mass region and (b) $\rho(770)$ region.}
\label{fig:effcosH}
\end{figure}

The detection efficiency is determined from a sample of two million $\Bdb\rightarrow \jpsi (\rightarrow \mu^+\mu^-)\pi^+\pi^-$ simulated events that are generated uniformly in phase space. Both $s_{12}$ and $s_{13}$ are centered at about $18.4\gev^2$. We model the detection efficiency using the symmetric dimensionless Dalitz plot observables
\begin{equation}
x= s_{12}/{\rm \gev}^2-18.4,~~~~{\rm and}~~~~  y=s_{13}/{\rm \gev}^2-18.4.
\end{equation}
These variables are related
to $s_{23}$ since
\begin{equation}
s_{12}+s_{13}+s_{23}=m^2_B+m^2_{\jpsi}+m^2_{\pi^+}+m^2_{\pi^-}.\label{conver}
\end{equation}

The acceptance in $\cos \theta_{\jpsi}$ is not uniform, but depends on $s_{23}$, as shown in Fig.~\ref{fig:effcosH}. If the efficiency was independent of $s_{23}$, then the curves would have the same shape. On the other hand,  no clear dependence on $s_{12}$ is seen. Thus the efficiency model can be expressed as
\begin{equation}
\varepsilon(s_{12}, s_{23}, \theta_{\jpsi})=\varepsilon_1(x,y)\times \varepsilon_2(s_{23}, \theta_{\jpsi}).\label{eq:eff}
\end{equation}
To study the $\cos \theta_{\jpsi}$ acceptance, we fit the $\cos \theta_{\jpsi}$ distributions from  simulation in 24 bins of $m^2(\pi^+\pi^-)$ with the function
\begin{equation}
\varepsilon_2(s_{23},\theta_{\jpsi})=\frac{1+ a\cos^2\theta_{\jpsi}}{2+2a/3},\label{eq:cosHacc}
\end{equation}
giving 24 values of $a$ as a function of $m^2(\pi^+\pi^-)$. The resultant distribution shown in Fig.~\ref{fig:cosHacc} can be described by an exponential function
\begin{equation}
a(s_{23})= \exp(a_1+a_2 s_{23}),
\end{equation}
with $a_1= -1.48\pm0.20$ and $a_2=(-1.45\pm0.33)\gev^{-2}$.

\begin{figure}[h]
\centering
\includegraphics[width =0.5\textwidth]{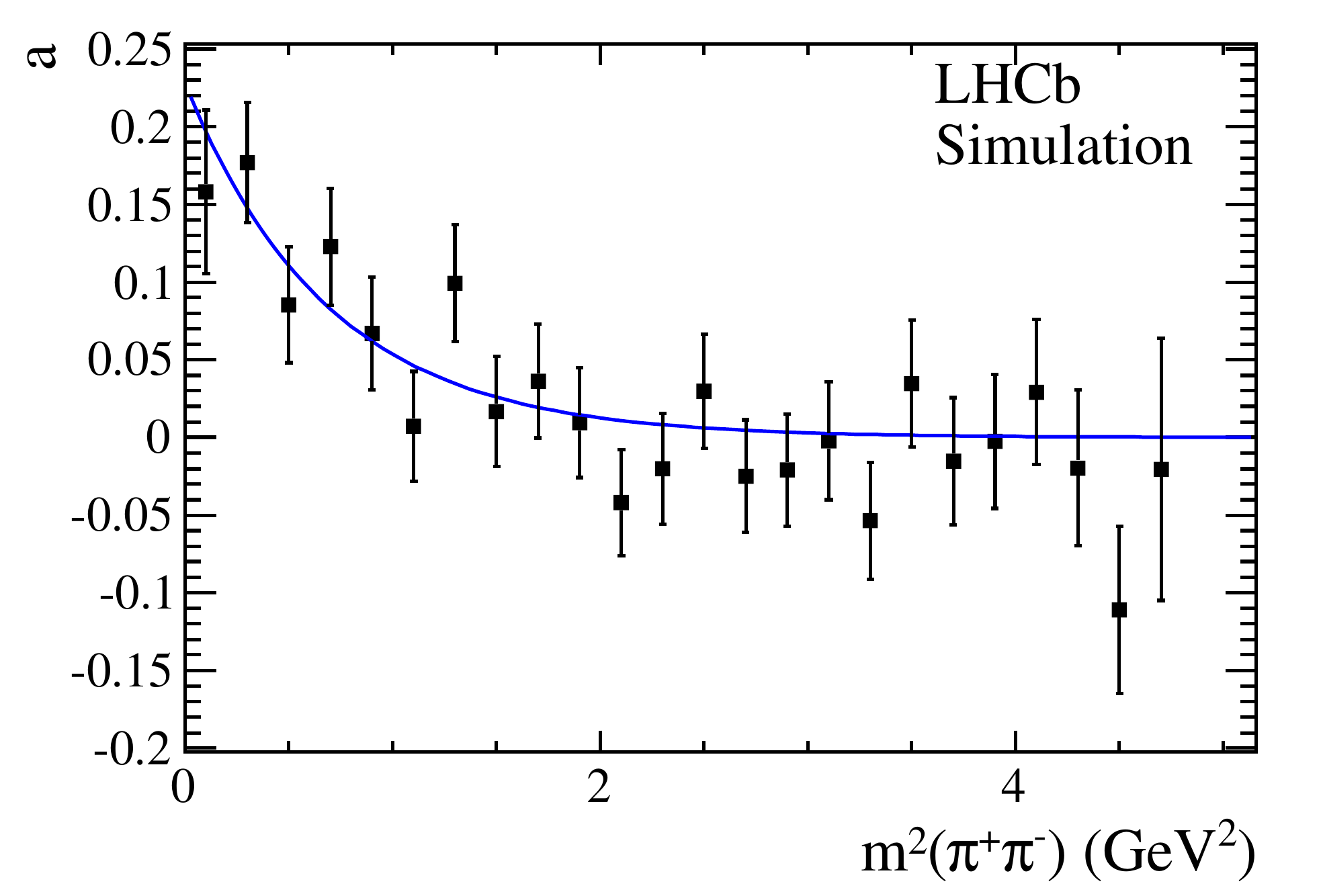}
\caption{Exponential fit to the acceptance parameter $a(s_{12})$ used in Eq.~\ref{eq:cosHacc}.}
\label{fig:cosHacc}
\end{figure}

Equation~\ref{eq:cosHacc} is normalized with respect to $\cos \theta_{\jpsi}$. Thus, after integrating over $\cos \theta_{\jpsi}$, Eq.~\ref{eq:eff} becomes
\begin{equation}
\int_{-1}^{+1}\varepsilon(s_{12}, s_{23}, \theta_{\jpsi})d\cos \theta_{\jpsi}=\varepsilon_1(x, y).
\end{equation}
This term of the efficiency is parametrized as a symmetric fourth order polynomial function given by
\begin{eqnarray}
\varepsilon_1(x,y)&=& 1+\epsilon'_1(x+y)+\epsilon'_2(x+y)^2+\epsilon'_3xy+\epsilon'_4(x+y)^3
+\epsilon'_5 xy(x+y)\nonumber \\
&&+\epsilon'_6(x+y)^4+\epsilon'_7 xy(x+y)^2+\epsilon'_8 x^2y^2,
\end{eqnarray}
where the $\epsilon'_i$ are the fit parameters.

\begin{table}[b]
\centering
\caption{Efficiency parameters to describe the acceptance on the signal Dalitz-plot.}
\begin{tabular}{lccc}
\hline
&Parameter& Value\\\hline
&$\epsilon'_1$ &~~$0.142\pm 0.010$ \\
&$\epsilon'_2$ &~~$0.101\pm 0.014$ \\
&$\epsilon'_3$ &~~$0.0082\pm 0.0005$ \\
&$\epsilon'_4$ &~~$0.027\pm 0.007$ \\
&$\epsilon'_5$ &~~$0.0052\pm 0.0003$ \\
&$\epsilon'_6$ &~~$0.0028\pm 0.0010$ \\
&$\epsilon'_7$ &~~$0.00074\pm 0.00017$ \\
&$\epsilon'_8$ &$-0.000105\pm 0.000008$ \\
\hline
&$\chi^2/ \rm ndf$ & 308/298\\
\hline
\end{tabular}
\label{tab:effparameter}
\end{table}

Figure~\ref{eff1} shows the polynomial function
\begin{figure}[t]
\begin{center}
     \includegraphics[scale=0.5]{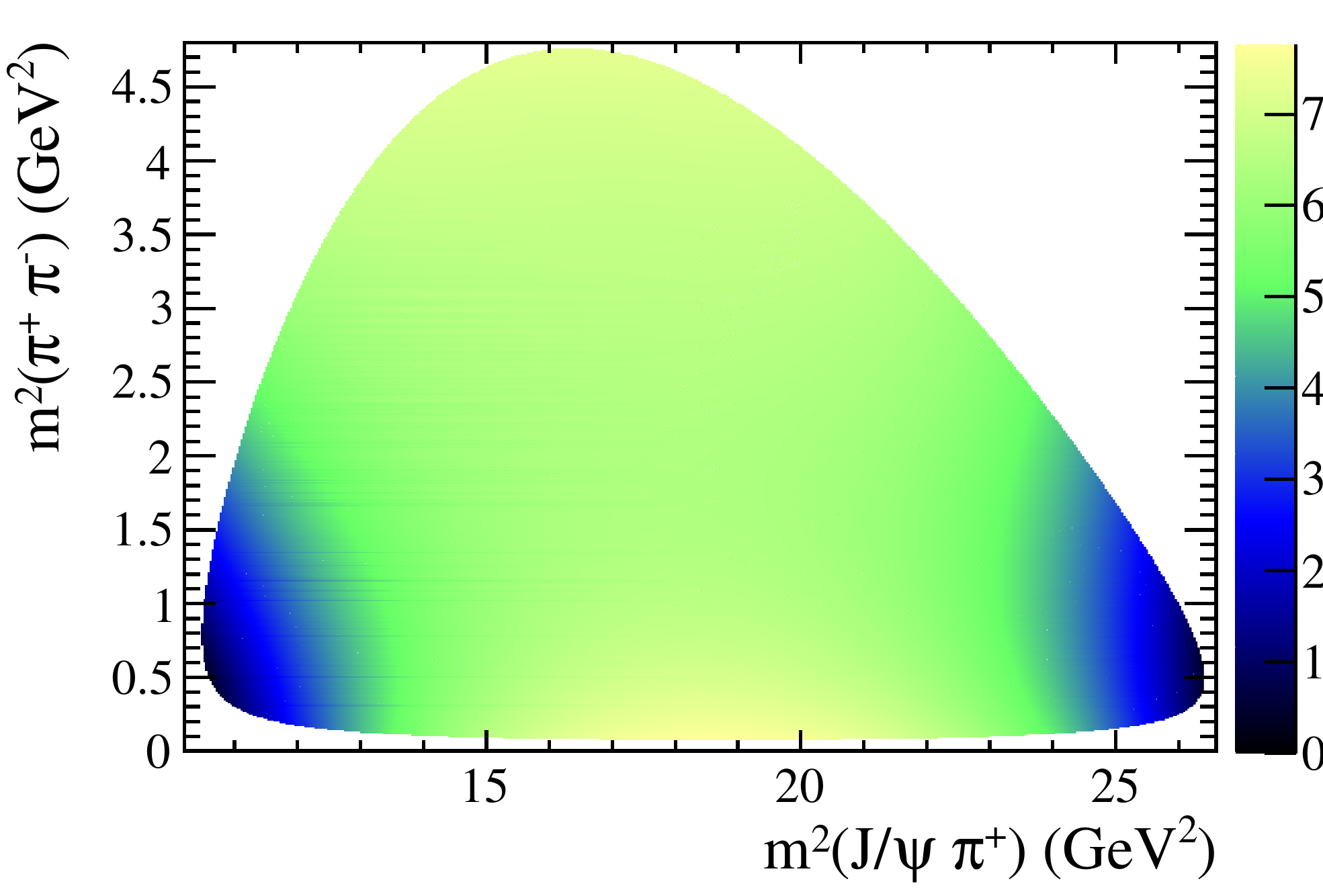}
\end{center}\label{eff1}
\vskip -0.5cm
\caption{Parametrized detection efficiency as a function
  of  $m^2(\pi^+\pi^-)$ versus $m^2(\jpsi\pi^+)$ determined from simulation. The $z$-axis scale is arbitrary.}
\end{figure}
obtained from a fit to the Dalitz-plot distributions of simulated events. The projections of the fit are shown in Fig.~\ref{eff2} and the resulting parameters are given in Table~\ref{tab:effparameter}. 
\begin{figure}[t]
\begin{center}
    \includegraphics[width=0.48\textwidth]{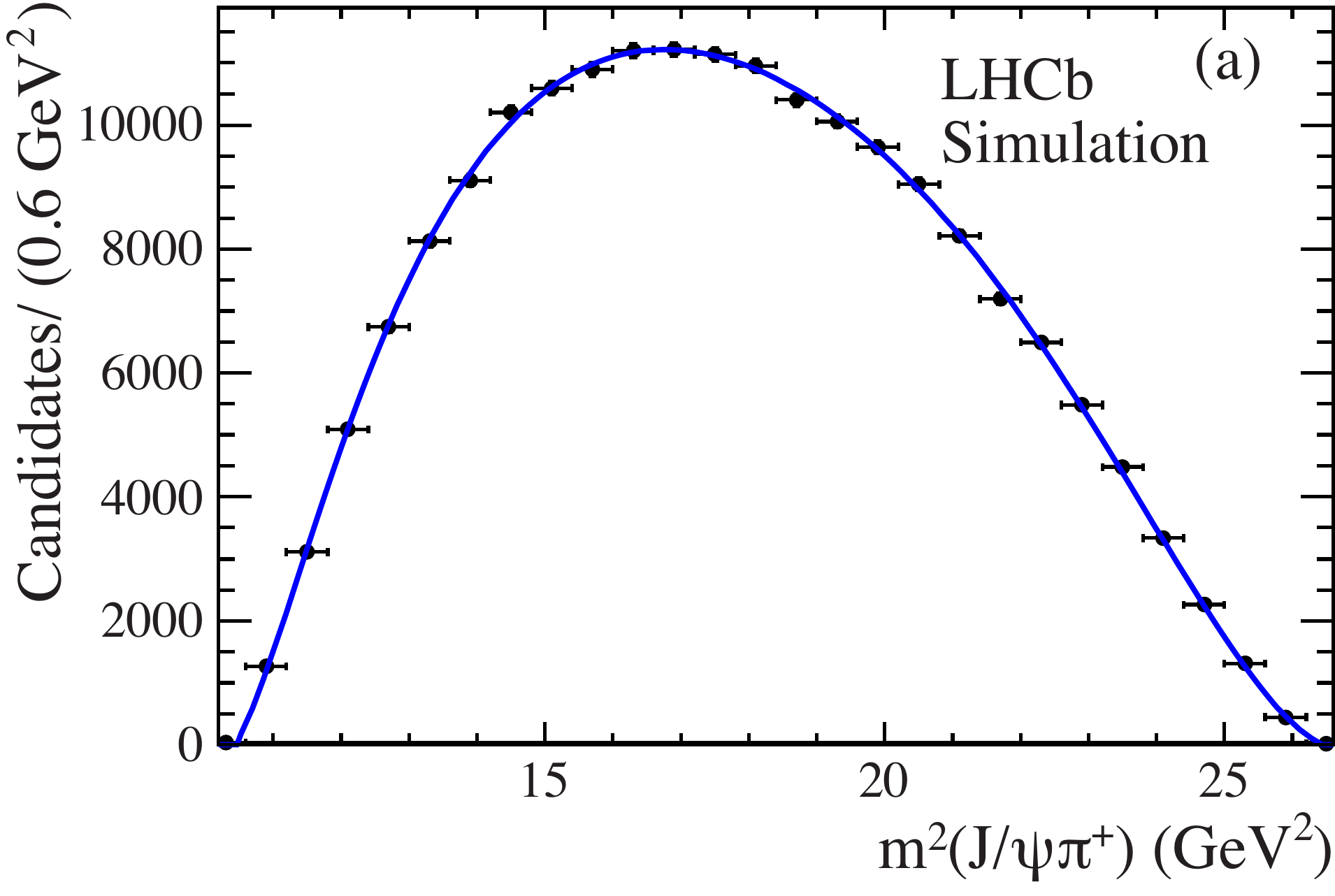}%
    \includegraphics[width =0.48\textwidth]{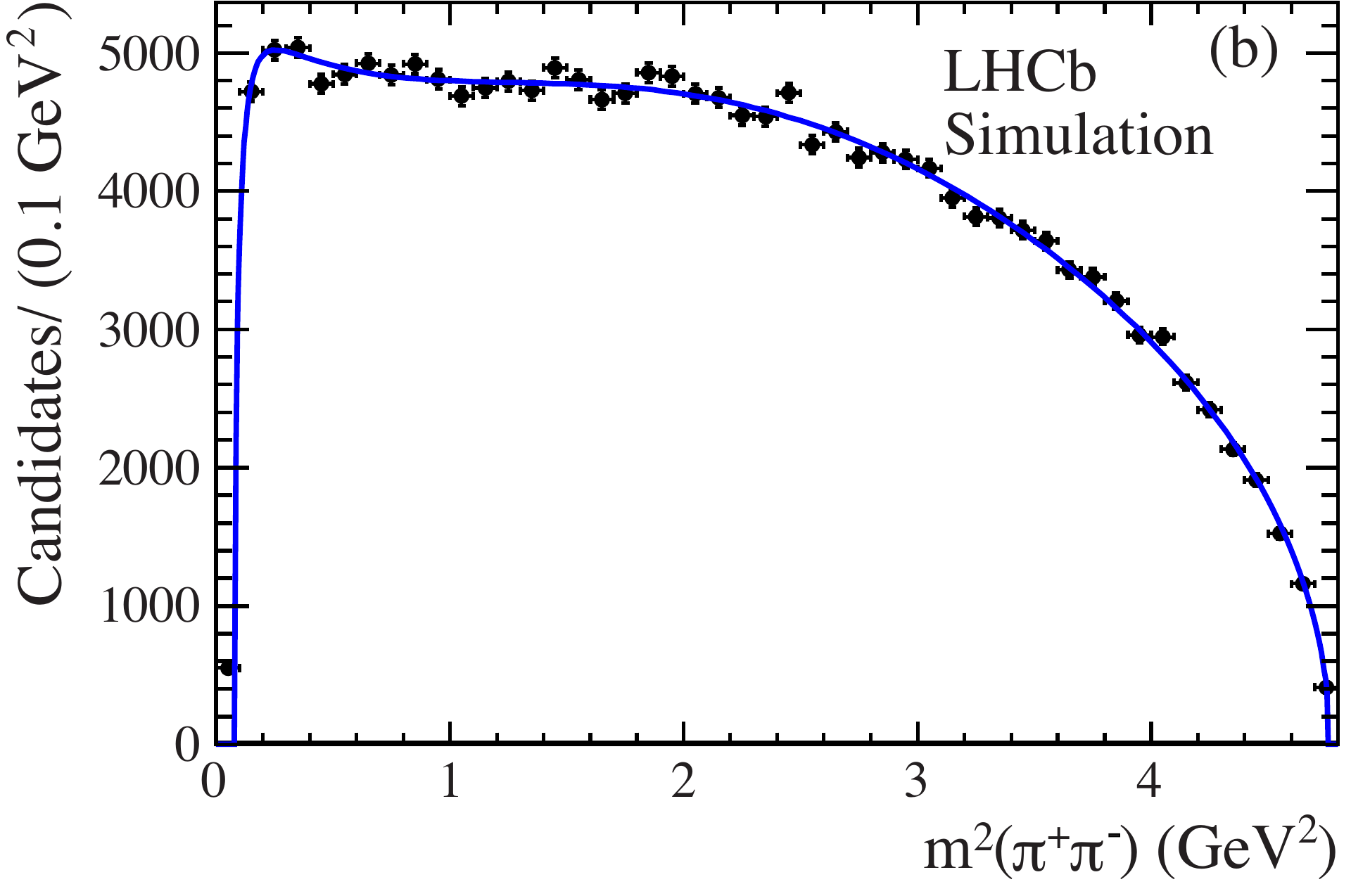}
\end{center}\label{eff2}
\vskip -0.5cm
\caption{Projections onto (a) $m^2(\jpsi \pi^+)$ and (b) $m^2(\pi^+\pi^-)$ of the simulated Dalitz plot used to determine the efficiency parameters. The points represent the simulated event distributions and the curves the projections of the polynomial fits.}
\end{figure}

\subsubsection{Background composition}
Backgrounds from $B$ decays into \jpsi final states have already been discussed in Section~\ref{sec:selections}. The main background source is combinatorial and its shape can be determined from
the same-sign $\pi^{\pm}\pi^{\pm}$ combinations within $\pm20$\mev of the $\Bdb$ mass peak; this region also contains the small $B^-$ background. In addition, there is background arising from partially reconstructed $\Bsb$ decays including $\Bsb\rightarrow \jpsi\eta' (\rightarrow \rho \gamma)$,
$\Bsb\rightarrow \jpsi\phi (\rightarrow \pi^+\pi^-\pi^0)$, and a $\Bdb\rightarrow \jpsi K^- \pi^+$ reflection, which cannot be present in same-sign combinations. We use simulated samples of inclusive $\Bsb$ decays, and exclusive $\Bdb\to \jpsi \Kbar^{*0}(892)$ and $\Bdb\to \jpsi \Kbar^{*0}_2(1430)$ decays to model the additional backgrounds.  The background fraction of each source is studied by fitting the $\jpsi \pi^+\pi^-$ candidate invariant mass distributions in bins of $m^2(\pi^+\pi^-)$. The resulting background distribution in the $\pm20$\mev $\Bdb$ signal region is shown in Fig.~\ref{bkgcmp}. It is fit by histograms from the same-sign combinations and two additional simulations, giving a partially reconstructed $\Bsb$ background of 12.8\%, and a reflection background that is 5.2\% of the total background.
\begin{figure}[htb]
\begin{center}
    \includegraphics[width=4 in]{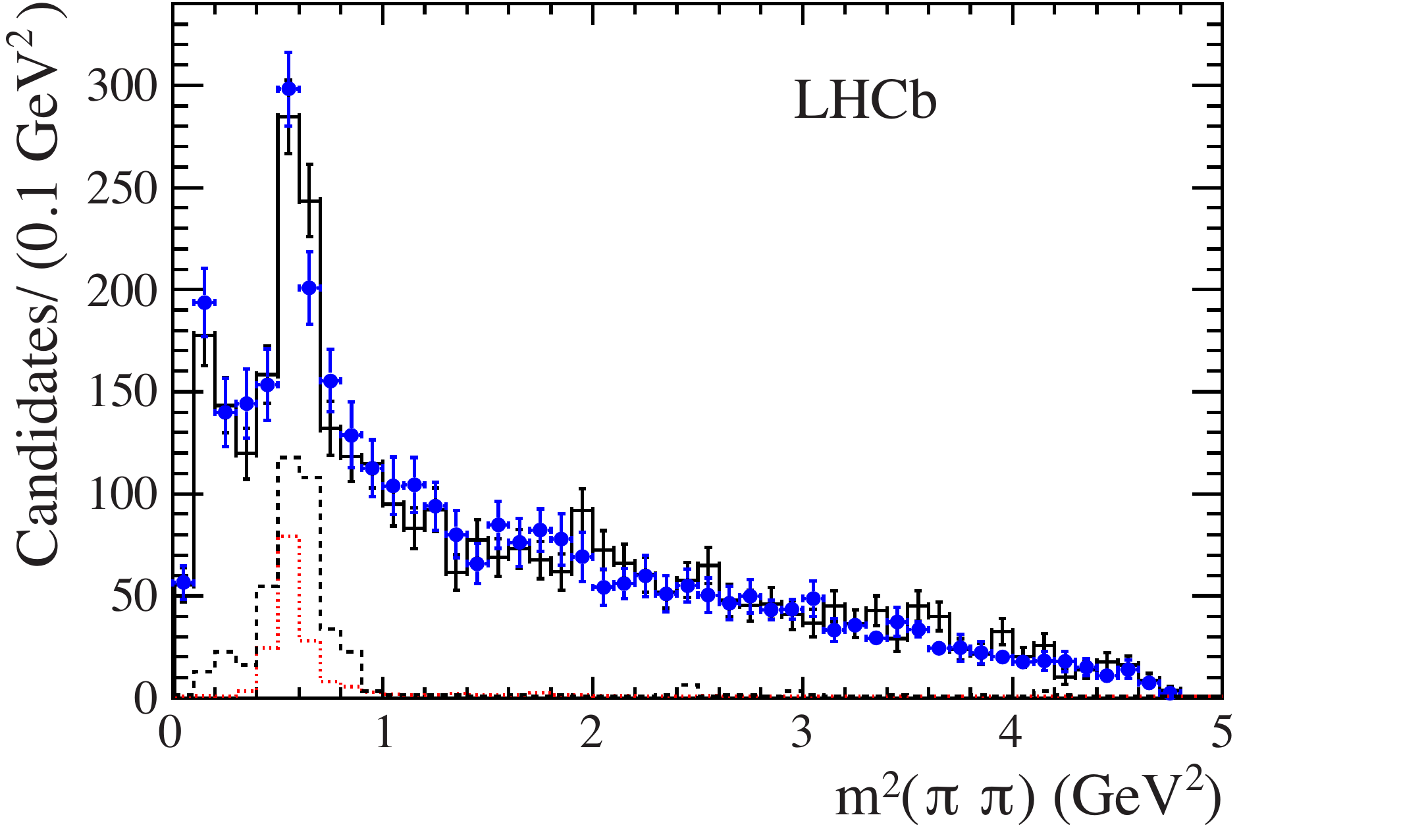}
\end{center}\label{bkgcmp}
\vskip -0.9cm
\caption{The $m^2(\pi\pi)$ distribution of background. The (black) histogram with error bars shows the same-sign data combinations with additional background from simulation, the (blue) points with error bars show the background obtained from the mass fits, the (black) dashed line is the partially reconstructed $\Bsb$ background, and the (red) dotted is the misidentified $\Bdb\rightarrow \jpsi K^- \pi^+$ contribution.}
\end{figure}

The background is parametrized as
\begin{equation}
B(s_{12}, s_{23}, \theta_{\jpsi})=\frac{m(\pi^+\pi^-)}{2P_R P_B m_{B}}B_1(s_{23},\cos \theta_{\pi\pi})\times \left(1+\alpha\cos^2\theta_{\jpsi}\right),
\end{equation}
where the first part $\frac{m(\pi^+\pi^-)}{2P_R P_B m_{B}}$ converts phase space from $s_{12}$ to $\cos \theta_{\pi\pi}$, and
\begin{eqnarray}\label{eq:bkgdlz}
B_1(s_{23},\cos \theta_{\pi\pi})=&&\left[B_2(\zeta)\frac{p_B}{m_B}+\frac{b_0}{(m^2_0-s_{23})^2+m_0^2\Gamma_0^2}\right] \nonumber\\
&&\times\frac{1+c_1q(\zeta)|\cos\theta_{\pi\pi}| +c_4p(\zeta)\cos^2 \theta_{\pi\pi}}{2[1+c_1q(\zeta)/2+c_4p(\zeta)/3]}.
\end{eqnarray}
The variable $\zeta=2(s_{23}-s_{\rm min})/(s_{\rm max}-s_{\rm min})-1$, where $s_{\rm min}$ and $s_{\rm max}$ give the fit boundaries,   $B_2(\zeta)$ is a fifth-order Chebychev polynomial with parameters $b_i$ ($i=1$--5), and $q(\zeta)$ and $p(\zeta)$ are both second-order Chebychev polynomials with parameters $c_i$ ($i$=2, 3, 5, 6), and $c_1$, and $c_4$ are free parameters. In order to better approximate the real background in the \Bsb signal region, the $\jpsi\pi^{\pm}\pi^{\mp}$ candidates are kinematically constrained to the \Bsb mass.  A fit to the same-sign sample,  with additional background from simulation, determines $b_i$, $c_i$, $m_0$ and $\Gamma_0$. 
Figure~\ref{bkg2} shows the mass squared projections from the fit.  The fitted background parameters are shown in Table \ref{tab:bkgparameter}.

The $\left(1+\alpha \cos^{2}\theta_{\jpsi}\right)$ term is a function of the $\jpsi$ helicity angle. The $\cos\theta_{\jpsi}$ distribution of background is shown in Fig.~\ref{bkg3}, and is fit with the function $1+\alpha\cos^2\theta_{\jpsi}$ that determines the parameter $\alpha=-0.38\pm0.04$. We have verified that $\alpha$ is independent of $s_{23}$.

\begin{figure}[htb]
\begin{center}
    \includegraphics[width=3. in]{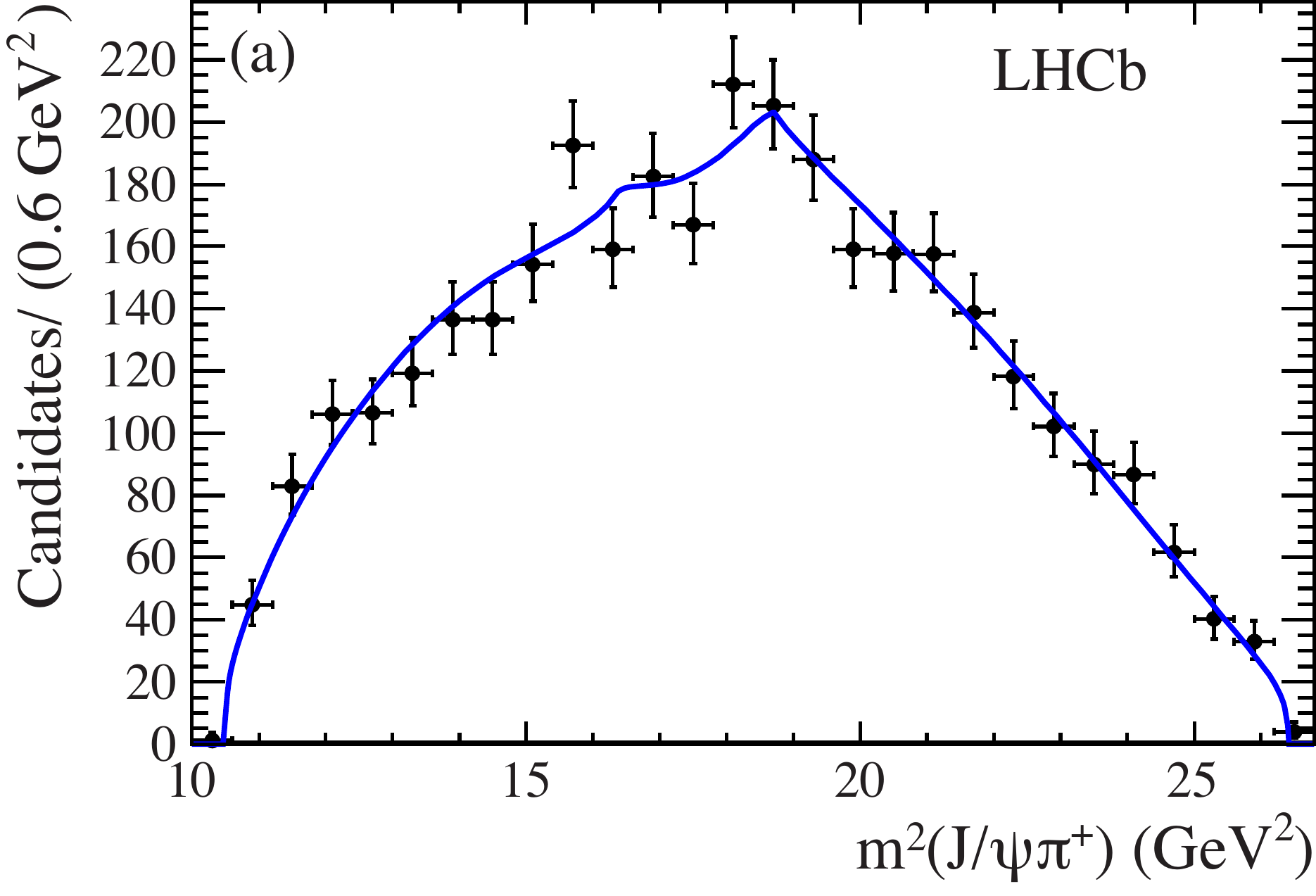}%
    \includegraphics[width=3.04 in]{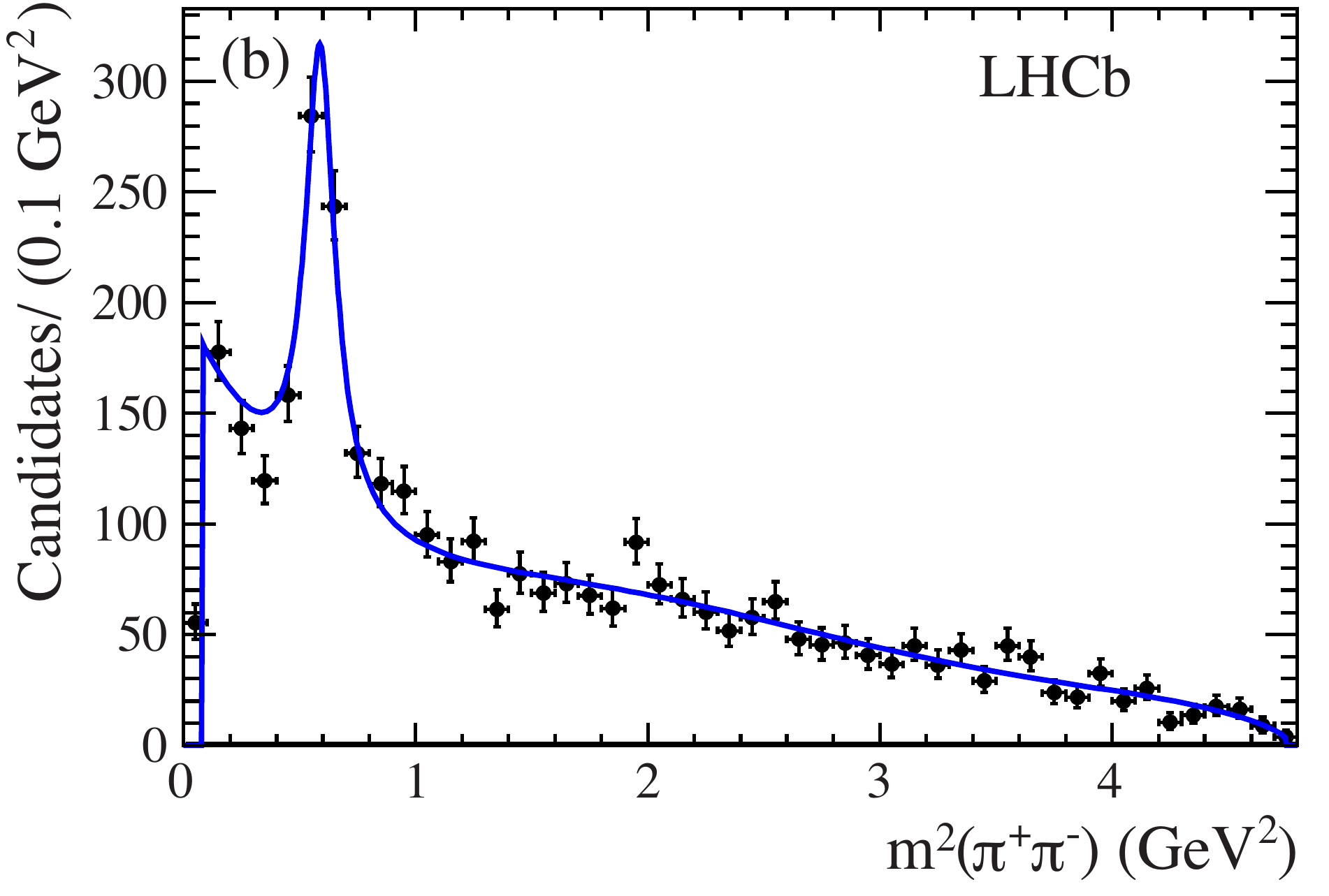}
\end{center}\label{bkg2}
\vskip -0.5cm
\caption{Projections of invariant mass squared of (a) $m^2(\jpsi \pi^+)$ and (b) $m^2(\pi^+\pi^-)$ of the background Dalitz  plot. The points with error bars show the same-sign combinations with additional background from simulation.}
\end{figure}
\begin{figure}[htb]
\vskip -6mm
\begin{center}
\includegraphics[scale=0.4]{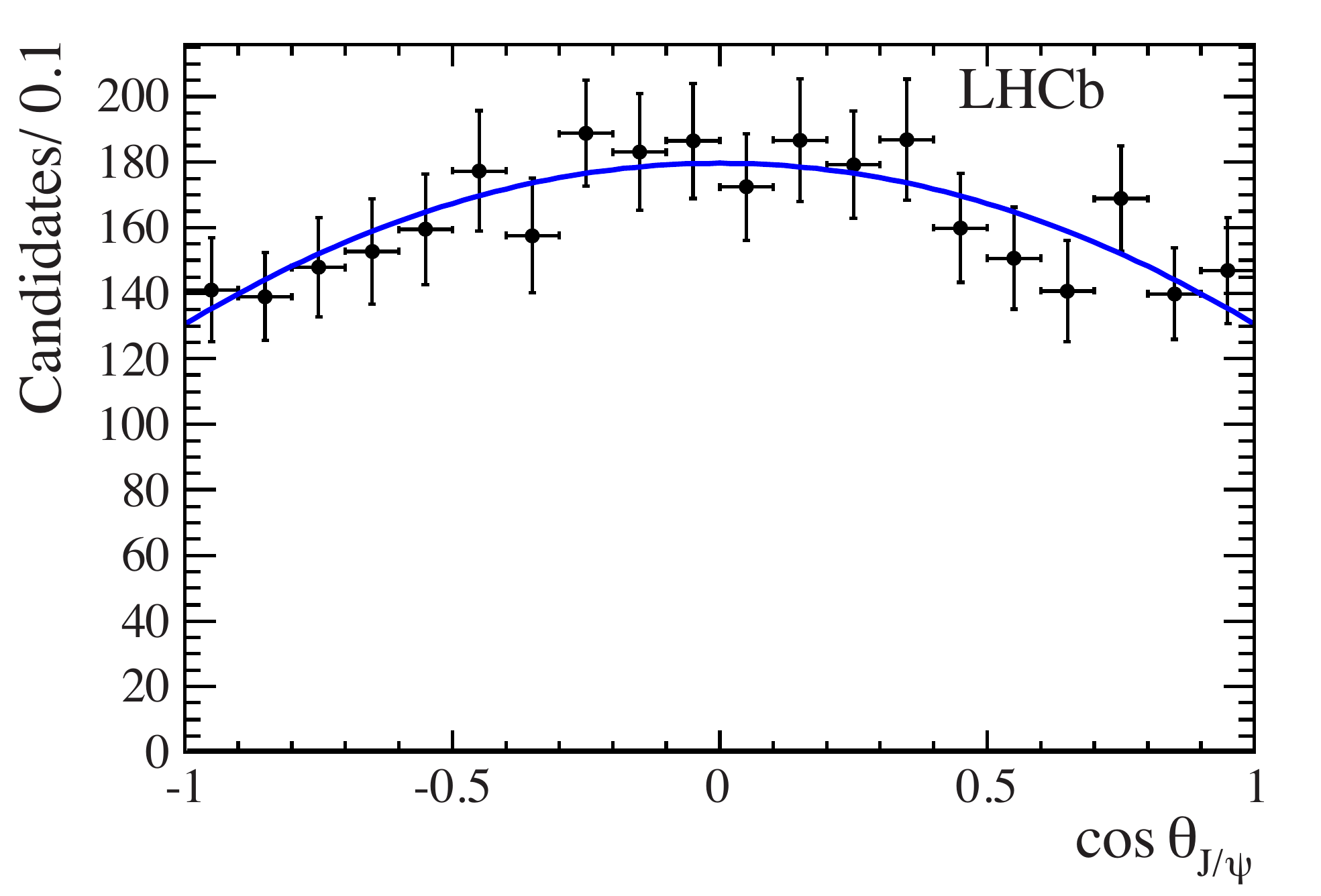}
\end{center}\label{bkg3}
\vskip -1cm
\caption{distribution of the background in $\cos\theta_{\jpsi}$ resulting from $\jpsi \pi^+\pi^-$ candidate mass fits in each bin of $\cos\theta_{\jpsi}$. The curve represents the fitted function $1+\alpha\cos^2\theta_{\jpsi}$.}
\end{figure}

\begin{table}[h!t!p!]
\centering
\caption{Parameters for the background model used in Eq.~\ref{eq:bkgdlz}.}
\begin{tabular}{lccc}
\hline
&Parameter & Value\\\hline
&$b_0$ &$(4.4\pm1.2)\times10^{-3}\gev^{4}$\\
&$m_0$ &~~~~~~~\!$0.767\pm 0.005\gev$\\
&$\Gamma_0$ &~~~~~~~\!$0.101\pm0.015 \gev$\\
&$b_1$ &$-0.52\pm 0.07$~~~\\
&$b_2$ &$0.22\pm 0.05$ \\
&$b_3$ &$-0.14\pm 0.06$~~~\\
&$b_4$ &$0.11\pm 0.04$\\
&$b_5$ &$-0.06\pm 0.04$~~~\\
&$c_1$ &$-0.70\pm 0.04$~~~\\
&$c_2$ &$-0.4\pm 0.3$~~~\\
&$c_3$ &$1.9\pm 0.2$\\
&$c_4$ &$0.42\pm 0.03$\\
&$c_5$ &$1.7\pm 0.8$\\
&$c_6$ &$2.5\pm 0.8$\\
\hline
&$\chi^2/{\rm ndf}$ & 252/284 \\
\hline
\end{tabular}
\label{tab:bkgparameter}
\end{table}

\subsection{Fit fractions}
While a complete description of the decay is given in terms of the fitted amplitudes and phases, the knowledge of the contribution of each component can be
summarized by defining a fit fraction, ${\cal{F}}^R_{\lambda}$, as the integration of the squared amplitude of $R$ over the Dalitz plot divided by the integration of the entire signal function,
\begin{equation}\label{eq:ff}
{\cal{F}}^R_{\lambda}=\frac{{ \int}\left| a^R_{\lambda} e^{i\phi^R_{\lambda}} \mathcal{A}_{\lambda}^{R}(s_{12},s_{23},\theta_{\jpsi})\right|^2 ds_{12}\;ds_{23}\;d\cos\theta_{\jpsi}}{{ \int} S(s_{12},s_{23},\theta_{\jpsi})  ~ds_{12}\;ds_{23}\;d\cos\theta_{\jpsi}}.
\end{equation}
Note that the sum of the fit fractions over all $\lambda$ and $R$ is not necessarily unity due to the potential presence of interference between two resonances. If the Dalitz plot has more destructive interference than constructive interference, the total fit fraction will be greater than one. Interference term fractions are given by
\begin{equation}
\label{eq:inter}
{\cal{F}}^{RR'}_{\lambda}=\mathcal{R}e\left(\frac{{ \int} a^R_{\lambda}\; a^{R'}_{\lambda} e^{i(\phi^R_{\lambda}-\phi^{R'}_{\lambda})} \mathcal{A}_{\lambda}^{R}(s_{12},s_{23},\theta_{\jpsi}) {\mathcal{A}_{\lambda}^{R'}}^{*}(s_{12},s_{23},\theta_{\jpsi}) ds_{12}\;ds_{23}\;d\cos\theta_{\jpsi}}{{ \int} S (s_{12},s_{23},\theta_{\jpsi}) ~ds_{12}\;ds_{23}\;d\cos\theta_{\jpsi}}\right),
\end{equation}
and the sum of the two is
\begin{equation}
\sum_{\lambda}\left(\sum_R {\cal{F}}^R_{\lambda}+\sum_{RR'}^{R\neq R'} {\cal{F}}^{RR'}_{\lambda}\right) =1.
\end{equation}
Note  that interference terms between different spin-$J$ states vanish, because the $d^J_{\lambda0}$ angular functions in $\mathcal{A}^R_{\lambda}$ are orthogonal.

The statistical errors of the fit fractions depend on the statistical errors of every fitted magnitude and phase, and their correlations. Therefore, to determine the uncertainties the covariance matrix and parameter values from the fit are used to generate 500 sample parameter sets. For each set, the fit fractions are calculated. The distributions of the obtained fit fractions are described by bifurcated Gaussian functions. The widths of the Gaussians are taken as the statistical errors on the corresponding parameters.
The correlations of fitted parameters are also taken into account.
\section{Final state composition}\label{fit}
\label{sec:Results1}
\subsection{Resonance models}

To study the resonant structures of the decay $\Bdb\rightarrow \jpsi \pi^+\pi^-$ 
we use those combinations with an invariant mass within
$\pm20$\mev of the $\Bdb$  mass peak and apply a $ \jpsi \KS$ veto. The total
number of remaining candidates is 8483, of which  $3212\pm80$ are attributed
to background.
Possible resonances  in the decay $\Bdb\rightarrow \jpsi \pi^+\pi^-$ are listed in Table \ref{reso1}. 
In addition, there could be some contribution from non-resonant $\Bdb\rightarrow \jpsi \pi^+\pi^-$ decays.

\begin{table}[hb]
\vspace*{-3mm}
\centering
\caption{Possible resonances in the $\Bdb\rightarrow \jpsi \pi^+\pi^-$ decay mode.}
\begin{tabular}{cccc}
\hline
Resonance & Spin & Helicity & Resonance \\
&&& formalism \\
\hline
$f_0(500)$  & 0 & 0 & BW \\
$\rho(770)$ & 1 & $0,\pm 1$ & BW \\
$\omega(782)$& 1 & $0,\pm 1$ & BW \\
$f_0(980)$ & 0 & 0 & Flatt\'e \\
$f_2(1270)$ & 2 &  $0,\pm 1$ & BW \\
$f_0(1370)$ & 0 & 0 & BW \\
$\rho(1450)$ & 1 & $0,\pm 1$ & BW \\
$f_0(1500)$ & 0 & 0 & BW \\
$\rho(1700)$ & 1 & $0,\pm 1$ & BW \\
$f_0(1710)$ & 0 & 0 & BW \\
\hline
\end{tabular}\label{reso1}
\end{table}

\begin{table}[ht]
\label{tab:resparam}
\vspace*{-3mm}
\centering
\caption{Breit-Wigner resonance parameters.}\label{PDG_param}
\begin{tabular}{cccc}
\hline
 Resonance &Mass (\mev) & Width (\mev) & Source \\
 \hline
 $f_0(500)$~~& ~~\,$513\pm32$ & ~~\!$335\pm67$ & CLEO \cite{Muramatsu:2002jp}\\
$\rho(770)$~ & ~$775.49\pm0.34$ &$149.1\pm0.8$&PDG \cite{Beringer:2012}\\
$\omega(782)$~ & ~$782.65\pm 0.12$ & ~~~$8.49\pm 0.08$ &PDG \cite{Beringer:2012}\\
$f_2(1270)$ & \!$1275.1\pm1.2$ & $185.1_{-2.4}^{+2.9}$&PDG \cite{Beringer:2012} \\
$f_0(1370)$ &  \!$1475\pm 6$& ~~$113\pm 11$ &\lhcb \cite{LHCb:2012ae}\\
$\rho(1450)$ & ~$1465\pm 25$ & ~~$400\pm 60$ &PDG \cite{Beringer:2012}\\
$f_0(1500)$ & \!$1505\pm6$ & ~$109\pm 7$&PDG \cite{Beringer:2012}\\
$\rho(1700)$ & ~$1700\pm 20$ & ~~~~$250\pm 100$ &PDG \cite{Beringer:2012}\\
$f_0(1710)$ & \!$1720\pm 6$ & ~$135\pm 8$ &PDG \cite{Beringer:2012}\\
\hline
\end{tabular}
\end{table}

The masses and widths of the BW resonances are listed in Table~\ref{PDG_param}. When used in the fit they are fixed to these values except for the parameters of the $f_0(500)$ resonance which are constrained by their uncertainties.
Besides the mass and width, the Flatt\'e resonance shape has two additional parameters $g_{\pi\pi}$ and $g_{KK}$, which are also fixed in the fit to values obtained in our previous Dalitz analysis of $\Bsb \to \jpsi\pip\pim$~\cite{LHCb:2012ae}, where a large fraction of $\Bsb$  decays are to $\jpsi f_0(980)$. The parameters are taken to be $m_0=939.9\pm 6.3$\mev, $g_{\pi\pi}=199\pm 30$\mev and $g_{KK}/g_{\pi\pi}=3.0\pm 0.3$.  All background and efficiency parameters are fixed in the fit.


To determine the complex amplitudes in a specific model, the data are fitted maximizing the unbinned likelihood given as
 \begin{equation}
\mathcal{L}=\prod_{i=1}^{N}F(s^i_{12},s^i_{23},\theta^i_{\jpsi}),
\end{equation}
where $N$ is the total number of candidates, and $F$ is the total PDF defined in Eq.~\ref{eq:pdf}. The PDF is constructed from the signal fraction $f_{\rm sig}$, the efficiency model $\varepsilon (s_{12}, s_{23}, \theta_{\jpsi})$, the background model $B(s_{12}, s_{23},\theta_{\jpsi})$, and the signal model $S(s_{12}, s_{23},\theta_{\jpsi})$. In order to ensure proper convergence using the maximum likelihood method, the PDF needs to be normalized. This is accomplished by first normalizing the $\jpsi$ helicity dependent part $\varepsilon(s_{23},\theta_{\jpsi}) \Theta_{\lambda}(\theta_{\jpsi})$ over $\cos \theta_{\jpsi}$ by analytical integration. This integration results in additional factors as a function of $s_{23}$. We then normalize the mass dependent part multiplied by the additional factors using numerical integration over 500$\times$500 bins.

The fit determines the relative amplitude magnitudes $a_\lambda^{R_i}$ and phases $\phi_\lambda^{R_i}$ defined in Eq.~\ref{amplitude-eq}; we choose to fix $a_0^{\rho(770)}$ to 1. As only relative phases are physically meaningful, one phase in each helicity grouping has to be fixed; we choose to fix those of  the $f_0(500)$ and the $\rho(770)$ ($|\lambda|=1$) to 0. In addition, since the final state $\jpsi \pi^+\pi^-$ is a self-charge-conjugate mode and as we do not determine the $B$ flavor, the signal function is an average of $\Bd$ and $\Bdb$ decays. If we do not consider $\pi^+\pi^-$ partial waves of a higher order than D-wave, then we can express the differential decay rate derived from Eqs. \ref{amplitude-eq}, \ref{eq4} and \ref{heli2} in terms of S-, P-, and D-waves including helicity 0 and $\pm1$
\begin{eqnarray}
&&\frac{d\overline\Gamma}{dm_{\pi\pi}d\cos\t d\cos\theta_{\jpsi}}\nonumber\\
&=& \left|\A^s_{S_0}e^{i\phi^s_{S_0}}+\A^s_{P_0} e^{i\phi^s_{P_0}}\cos \t+\A^s_{D_0} e^{i\phi^s_{D_0}} \left(\frac{3}{2}\cos^2 \t -\frac{1}{2}\right)\right|^2\sin^2\theta_{\jpsi}\nonumber\\
&+&\left|\A^s_{P_{\pm 1}} e^{i\phi^s_{P_{\pm 1}}} \frac{1}{2}\sin\t +\A^s_{D_{\pm 1}}e^{i\phi^s_{D_{\pm 1}}} \sqrt{\frac{3}{2}}\sin\t \cos\t\right|^2 \frac{1+\cos^2\theta_{\jpsi}}{2}\label{Rb}
\end{eqnarray}
for $\Bzb$ decays, where $\A^s_{k_\lambda}$ and $\phi^s_{k_\lambda}$ are the sum of amplitudes and reference phase for the spin-$k$ resonance group, respectively.
The $\Bd$ function for decays is similar, but $\theta_{\pipi}$ and $\theta_{\jpsi}$ are changed to $\pi-\theta_{\pipi}$ and $\pi-\theta_{\jpsi}$ respectively, as a result of using $\pi^-$ and $\mu^-$ to define the helicity angles, yielding
\begin{eqnarray}
&&\frac{d\Gamma}{dm_{\pi\pi}d\cos\t d\cos\theta_{\jpsi}}\nonumber\\&=& \left|\A^s_{S_0}e^{i\phi^s_{S_0}}-\A^s_{P_0} e^{i\phi^s_{P_0}}\cos \t+\A^s_{D_0} e^{i\phi^s_{D_0}}\left(\frac{3}{2}\cos^2 \t -\frac{1}{2}\right)\right|^2\sin^2\theta_{\jpsi}\nonumber\\
&+&\left|\A^s_{P_{\pm 1}} e^{i\phi^s_{P_{\pm 1}}} \frac{1}{2}\sin\t -\A^s_{D_{\pm 1}}e^{i\phi^s_{D_{\pm 1}}} \sqrt{\frac{3}{2}}\sin\t \cos\t\right|^2 \frac{1+\cos^2\theta_{\jpsi}}{2}.\label{R}
\end{eqnarray}
Summing Eqs. \ref{Rb} and \ref{R} results in cancellation of the interference involving the $\lambda=0$ terms for spin-1, and the $\lambda=\pm1$ terms for spin-2, as they appear with opposite signs for $\Bzb$ and $\Bz$ decays. Therefore, we have to fix one phase in spin-1 ($\lambda=0$) group ($\phi^s_{P_0}$) and one in spin-2 ($\lambda=\pm1$) group ($\phi^s_{D_{\pm 1}}$); the phases of $\rho(770)$ ($\lambda=0$) and $f_2(1270)$ ($\lambda=\pm1$) are fixed to zero. The other phases in each corresponding group are relative to that of the fixed resonance.

\subsection{Fit results}

To find the best model, we proceed by fitting with all the possible resonances and a non-resonance (NR) component, then subsequently remove the most insignificant component one at a time. We repeat this procedure until each remaining contribution has more than 3 statistical standard deviation ($\sigma$) significance. The significance is estimated from the fit fraction divided by its statistical uncertainty.  The best fit model contains six resonances, the $f_0(500)$, $f_0(980)$, $f_2(1270)$, $\rho(770)$, $\rho(1450)$, and $\omega(782)$. 

In order to compare the different models quantitatively an estimate of the goodness of fit is calculated from three-dimensional partitions of the one angular and two mass squared variables. We use the Poisson likelihood $\chi^2$ \cite{Baker:1983tu} defined as
\begin{equation}
\chi^2=2\sum_{i=1}^{N_{\rm bin}}\left[  x_i-n_i+n_i \text{ln}\left(\frac{n_i}{x_i}\right)\right],
\end{equation}
where $n_i$ is the number of events in the three dimensional bin $i$ and $x_i$ is the expected number of events in that bin according to the fitted likelihood function. A total of 1021 bins ($N_{\rm bin}$) are used to calculate the $\chi^2$, based on the variables $m^2(\jpsi\pi^+)$,  $m^2(\pi^+\pi^-)$, and $\cos\theta_{\jpsi}$.  The $\chi^2/\text{ndf}$ and the negative of the logarithm of the likelihood, $\rm -ln\mathcal{L}$, of the fits are given in Table~\ref{RMchi2}; ndf is equal to $N_{\rm bin} -1 - N_{\rm par}$, where $N_{\rm par}$ is the number of fitting parameters.
 The difference between the best fit results and fits with one additional component is taken as a systematic uncertainty. Figure~\ref{RM4} shows the best fit model projections of $m^2(\pi^+\pi^-)$, $m^2(\jpsi \pi^{+})$, $\cos \theta_{\jpsi}$ and $m(\pi^+\pi^-)$. 
We calculate the fit fraction of each component using Eq. \ref{eq:ff}. For a P- or D-wave resonance, we report its total fit fraction by summing all the helicity components, and the fraction of the helicity $\lambda=0$ component. The results are listed in Table \ref{ff3}.  Systematic uncertainties will be discussed in Section \ref{sec:syst}. Two interesting ratios of fit fractions are ($0.93_{-0.22-0.23}^{+0.37+0.47}$)\% for $\omega(782)$ to $\rho(770)$, and ($9.5_{-3.4}^{+6.7}\pm3.0$)\% for $f_0(980)$ to $f_0(500)$.

The fit fractions of the interference terms are computed using Eq.~\ref{eq:inter} and listed in Table~\ref{tab:inter}. Table~\ref{tab:phase} shows the resonant phases from the best fit. For the systematic uncertainty study, Table \ref{tab:ff1} shows the fit fractions of components for the best model with one additional resonance.

\begin{table}[htb]
\caption{Values of $\chi^2/\text{ndf}$ and $\rm -ln\mathcal{L}$ of different resonance models.}
\centering
\begin{tabular}{lccc}
\hline
Resonance model & $\rm -ln\mathcal{L}$& $\chi^2/\text{ndf}$ & Probability (\%) \\
\hline
Best Model &35292 & 1058/1003 & 11.1\\
\hline
Best Model + $\rho(1700)$ &35284 &1045/ 999 &15.0\\
Best Model + NR &35284 &1058/1001&10.3\\
Best Model + $f_0(1370)$ &35285 &1047/1001&15.2\\
Best Model + $f_0(1500)$ &35287 &1049/1001&14.4\\
Best Model + $f_0(1710)$ &35289 &1052/1001&12.6\\
\hline
\end{tabular}
\label{RMchi2}
\end{table}

\begin{figure}[b]
  \begin{center}
     \includegraphics[width=0.5\textwidth]{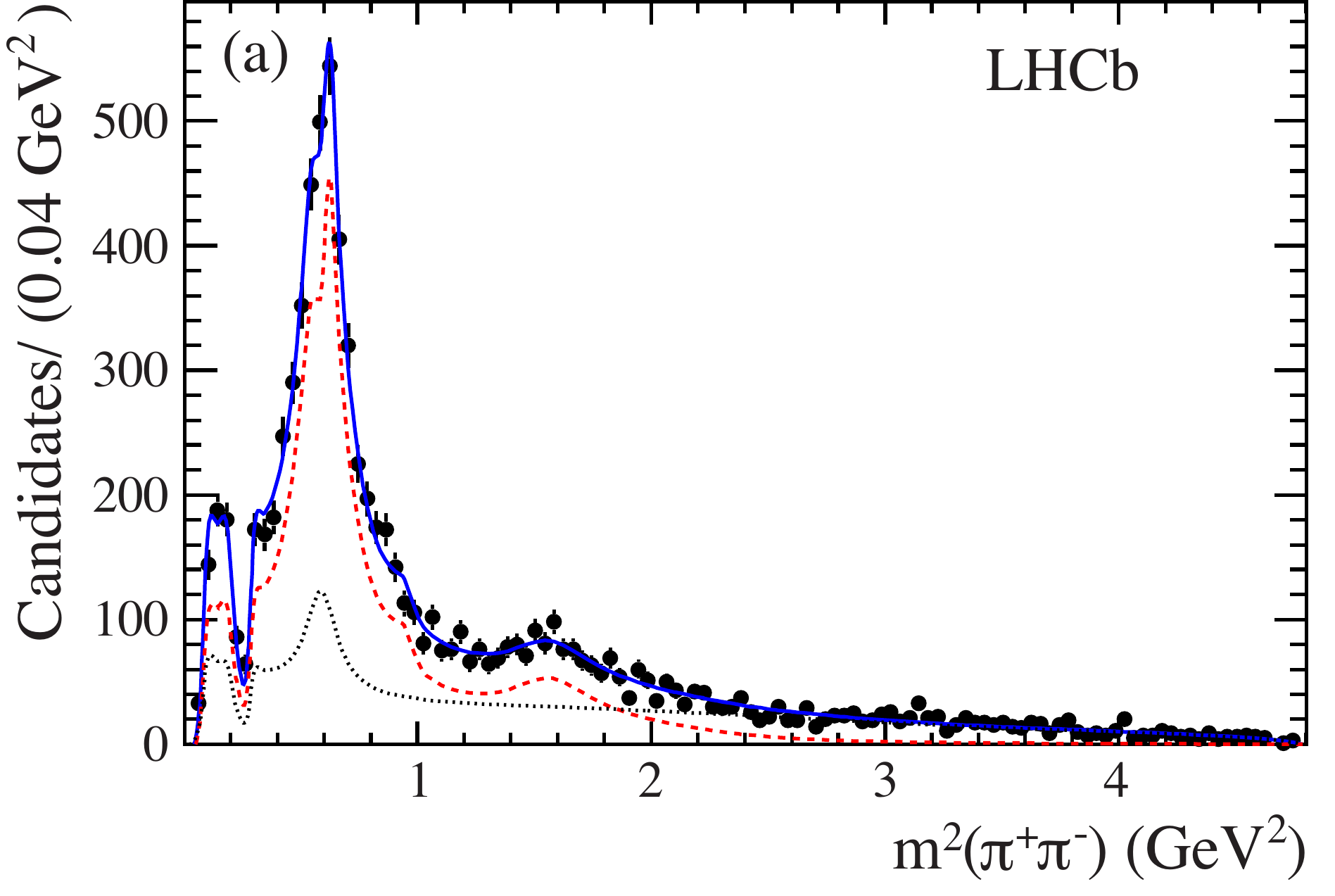}%
     \includegraphics[width=0.5\textwidth]{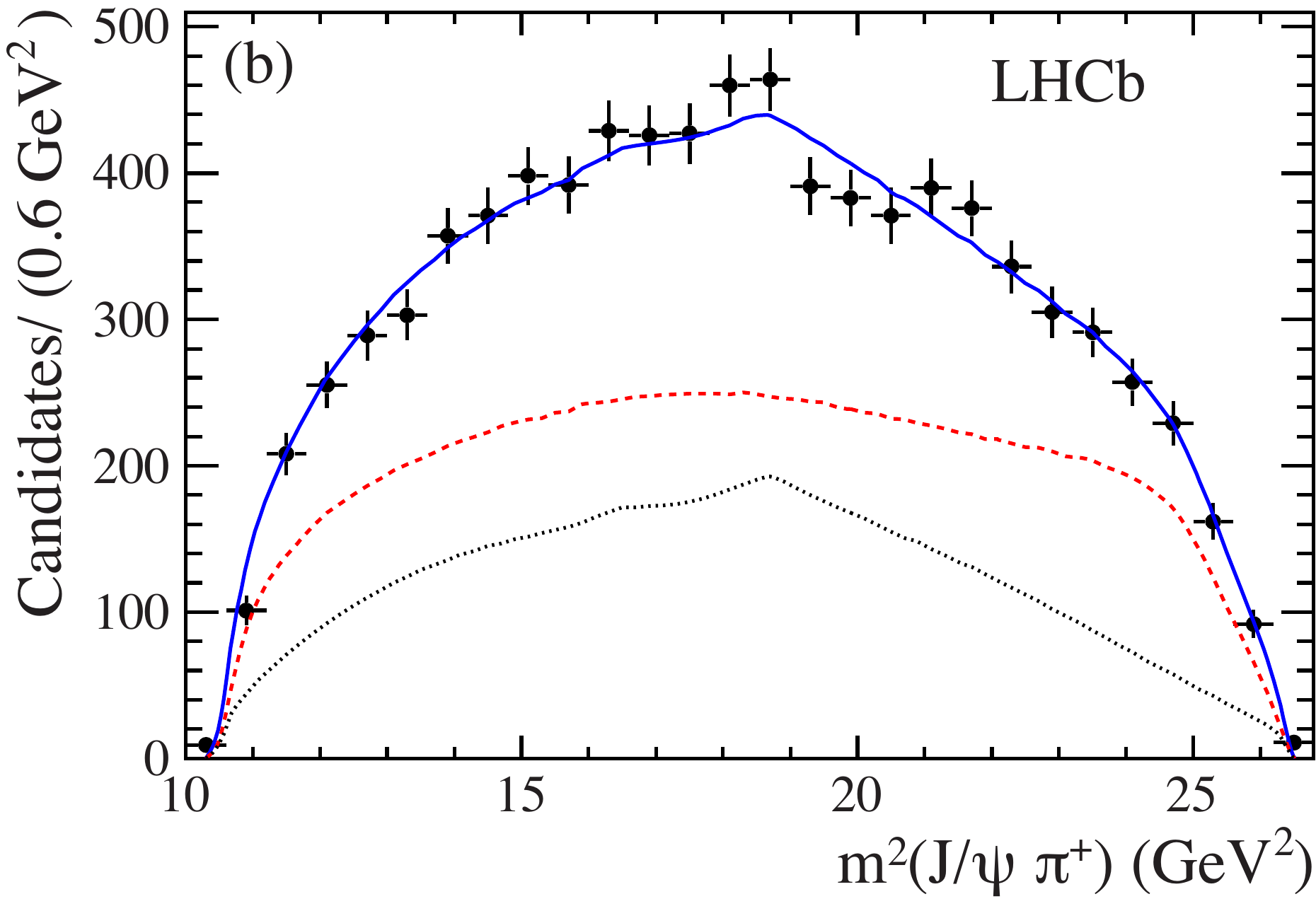}\\
      \includegraphics[width=0.5\textwidth]{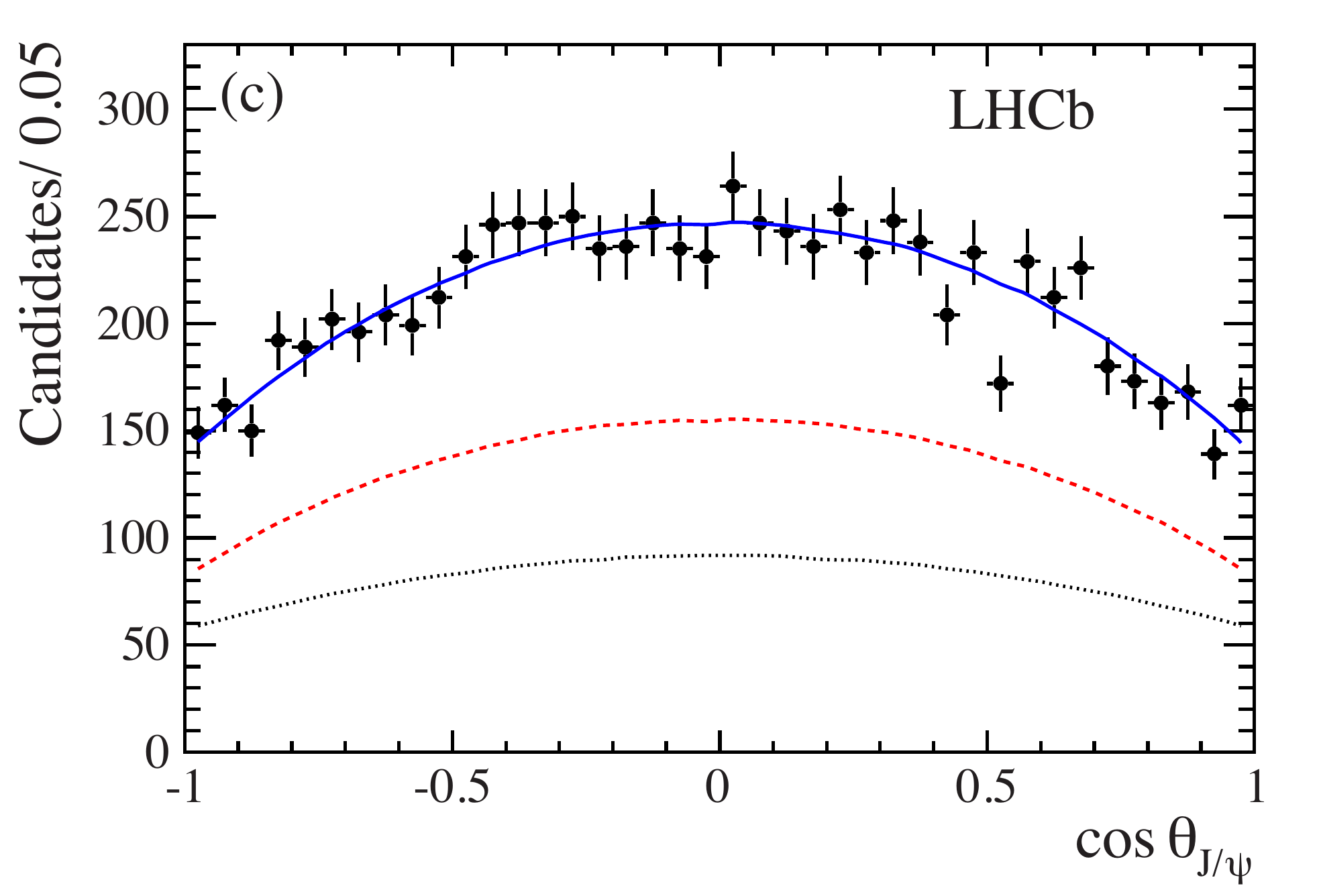}%
     \includegraphics[width=0.5\textwidth]{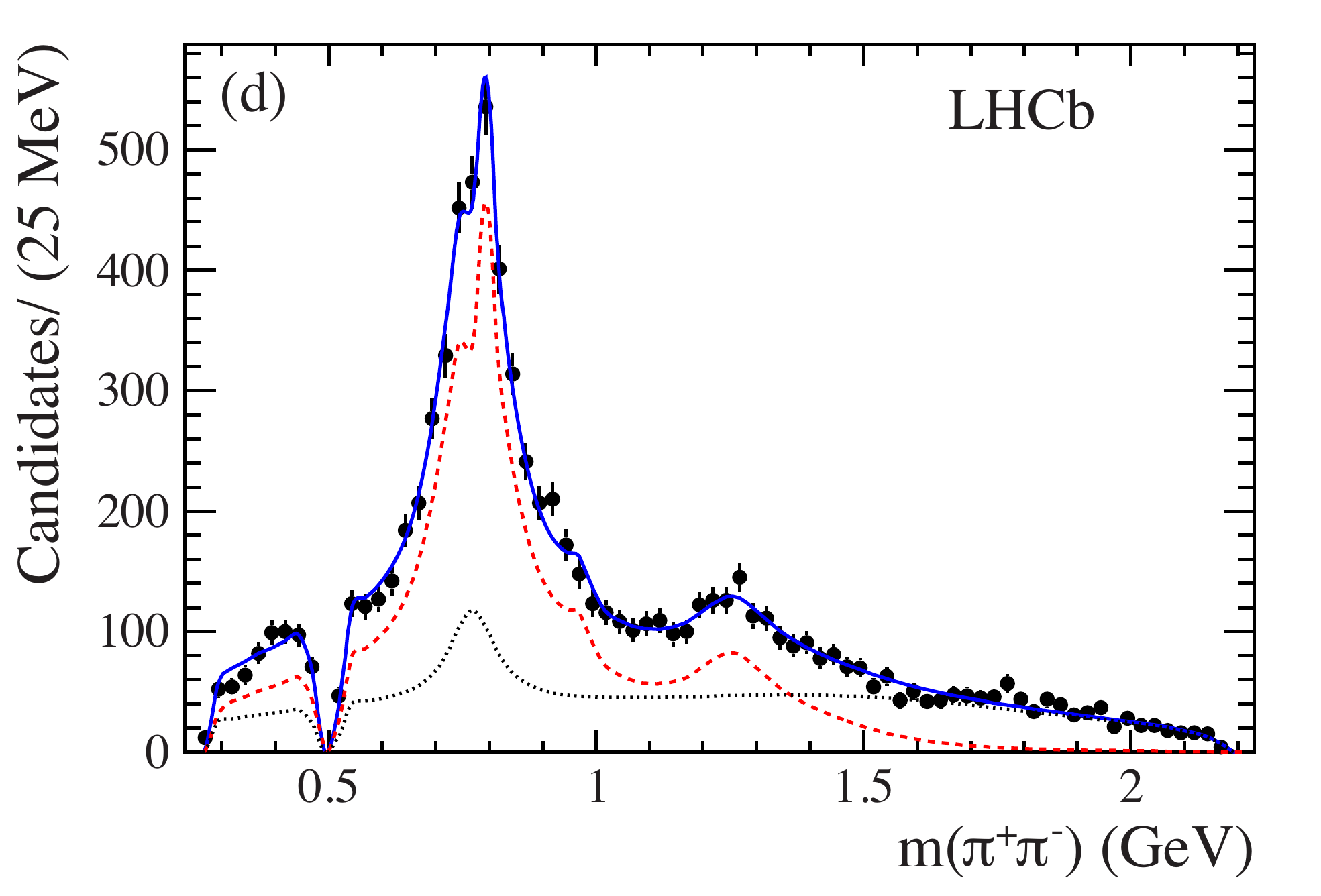}\\
    \caption{Dalitz fit projections of (a) $m^2(\pi^+\pi^-)$, (b) $m^2(\jpsi \pi^{+})$, (c) $\cos \theta_{\jpsi}$ and (d) $m(\pi^+\pi^-)$ for the best model. The points with error bars are data, the signal fit is shown with a (red) dashed line, the background with a (black) dotted line, and the (blue) solid line represents the total. In (a) and (d), the shape variations near the $\rho(770)$ mass is due to $\rho(770)-\omega(782)$ interference, and the dip at the $\KS$ mass~\cite{Beringer:2012} is due to the $\KS$ veto.}  \label{RM4}
  \end{center}
\end{figure}

\begin{table}[t]
\centering
\caption{Fit fractions and significances of contributing components for the best model, as well as the fractions of the helicity $\lambda=0$ part. The significance takes into account both statistical and systematic uncertainties.} 
\def\arraystretch{1.2}
\begin{tabular}{lccc}
\hline
Components & Fit fraction (\%)& $\lambda=0$ fraction & Significance ($\sigma$)\\
\hline
$\rho(770)$& $62.8_{-2.9-4.8}^{+4.8+2.8}$& $0.63\pm0.04_{-0.03}^{+0.06}$ & 11.2\\
$\omega(782)$& $0.59_{-0.13-0.14}^{+0.23+0.27}$ & $0.30^{+0.26}_{-0.18}\pm0.05$ &  3.1\\
$f_0(980)$& $1.53^{+0.77+0.43}_{-0.50-0.35}$ & 1 & 2.5\\
$f_2(1270)$& $8.9\pm1.1\pm1.0$& $0.76\pm0.06\pm0.05$ & 5.9\\
$\rho(1450)$& $5.3^{+2.5+5.6}_{-1.4-0.9}$& $0.28^{+0.17+0.08}_{-0.13-0.12}$ & 3.2\\
$f_0(500)$& $16.2\pm2.0^{+6.0}_{-2.0}$& 1 & 5.7\\
\hline
Sum&95.2&\\
\hline
\end{tabular}\label{ff3}
\end{table}

\begin{table}[ht]
\centering
\caption{Interference fractions ${\cal{F}}_{\lambda}^{RR'}$ (\%) computed using Eq \ref{eq:inter}. Note that the diagonal elements are fit fractions defined in Eq \ref{eq:ff}. }
\begin{tabular}{rc|cccccccccc}\hline
& &$\rho$ &$\omega$&$\rho$& $f_0$&$f_0$  &$f_2$ &$\rho$ &$\omega$&$\rho$&$f_2$ \\
& & 770   &  782   & 1450& 980 & 500 & 1270 & 770 & 782& 1450& 1270\\
&$|\lambda|$&0& 0 &0 & 0 & 0 & 0 & 1 &1 &1 & 1\\\hline
$\rho(770)$ &0& 39.44 & $-0.02$&$-0.89$&0&0&0&0&0&0&0\\
$\omega(782)$ &0& & ~~0.18& $-0.05$&0&0&0&0&0&0&0\\
$\rho(1450)$ &0& & & ~~1.47 &0&0&0&0&0&0&0\\
$f_0(980)$ & 0& & & & 1.53& ~~2.08& 0&0&0&0&0\\
$f_0(500)$ & 0 & & & & & 16.15&0&0&0&0&0\\
$f_2(1270)$ & 0 & & & & & & 6.72&0&0&0&0\\
$\rho(770)$ & 1& & & & & & & 23.32&0.29&0&0\\
$\omega(782)$ & 1& & & & & & & &0.41 &$-0.07$&0\\
$\rho(1450)$ &1& & & & & & & & & ~~3.80&0\\
$f_2(1270)$ &1&& & & & & & & & & 2.14\\\hline
\end{tabular}
\label{tab:inter}
\end{table}

\begin{table}[t]
\centering
\caption{Resonant phases from the best fit.}
\begin{tabular}{lr}
\hline
Components & Phase (deg) \\
\hline

$\rho(770)$, $\lambda =0$& 0 (fixed)\\
$\rho(770)$, $|\lambda| =1$& 0 (fixed)\\
$\omega(782)$, $\lambda =0$& $-84\pm31$\\
$\omega(782)$, $|\lambda| =1$& $-70\pm16$\\
$f_0(980)$& $103\pm17$\\
$f_2(1270)$, $\lambda =0$& $-87\pm12$\\
$f_2(1270)$, $|\lambda| =1$& 0 (fixed)\\
$\rho(1450)$, $\lambda =0$& $-162\pm22$\\
$\rho(1450)$, $|\lambda| =1$& $160\pm48$\\
$f_0(500)$& 0 (fixed)\\
\hline
\end{tabular}\label{tab:phase}
\end{table}

\begin{table}[hbtp]
\centering
\caption{Fit fractions (\%) of contributing components for the best model with adding one additional resonance.}
\def\arraystretch{1.2}
\begin{tabular}{lcccccc}
\hline
             &Best                  &$+\rho(1700)$          & $+f_0(1370)$          & $+f_0(1500)$          & $+f_0(1710)$ &+NR\\\hline
$\rho(770)$  &$62.8_{-2.9}^{+4.8}$  & $59.5_{-2.8}^{+3.1}$   & $62.6_{-2.5}^{+3.9}$  & $62.4_{-2.5}^{+4.1}$  & $63.3_{-3.0}^{+5.6}$ & $63.4_{-2.7}^{+3.8}$\\
$\omega(782)$&$0.59_{-0.13}^{+0.23}$& $0.58_{-0.15}^{+0.22}$& $0.60_{-0.15}^{+0.26}$& $0.60_{-0.15}^{+0.25}$& $0.59_{-0.15}^{+0.25}$& $0.59_{-0.14}^{+0.25}$\\
$f_0(980)$   &$1.53_{-0.50}^{+0.77}$& $1.54_{-0.53}^{+0.75}$& $1.54_{-0.55}^{+0.76}$& $1.50_{-0.54}^{+0.78}$& $1.55_{-0.51}^{+0.76}$& $1.74_{-0.55}^{+0.80}$\\
$f_2(1270)$  &$8.9\pm1.1$           & $8.1\pm1.2$           & $8.8\pm1.1$           & $8.8\pm1.1$           & $8.9\pm1.1$     & $8.8\pm1.1$\\
$\rho(1450)$ &$5.3_{-1.4}^{+2.5}$   & $10.8_{-3.6}^{+5.4}$  & $4.7_{-1.1}^{+1.6}$   & $4.9_{-1.2}^{+1.9}$   & $5.7_{-2.3}^{+4.0}$  & $4.6_{-1.3}^{+2.1}$\\
$f_0(500)$   &$16.2\pm2.0$          & $15.6\pm1.9$          & $16.6\pm2.0$          & $16.9\pm2.1$          & $16.3\pm2.1$ & $21.9\pm3.8$\\
$\rho(1700)  $&-                    & $3.4_{-1.5}^{+2.7}$    & -            & -            & - &-                                  \\
$f_0(1370)$  &-            & -                              & $1.3_{-0.5}^{+0.8}$   & -            &-  &-\\
$f_0(1500)$  &-            &-             & -            &                            $1.0_{-0.4}^{+0.7}$ &-  &-\\
$f_0(1710)$  &-            &-             & -            & -            &$0.4_{-0.2}^{+0.4}$ &-\\
NR           &-            &-             & -            & -            & -            &$4.5_{-1.7}^{+2.7}$\\
\hline
Sum          &95.2         &99.6          &96.2          &96.0          &96.7           &105.5\\
\hline
\end{tabular}\label{tab:ff1}
\end{table}

\clearpage
\subsection{Helicity angle distributions}
\label{sec:hel-dist}
We show the helicity angle distributions in the $\rho(770)$ mass region defined within one full width of the $\rho(770)$  resonance (the width values are given in Table~\ref{tab:resparam}) in Fig. \ref{plot:hel}.
The $\cos \theta_{\jpsi}$ and $\cos \theta_{\pi\pi}$ background subtracted and efficiency corrected distributions for this mass region are presented in Fig. \ref{helii1_4}.  The distributions are in good agreement with the best fit model.

\begin{figure}[htb]
\begin{center}
    \includegraphics[width=0.48\textwidth]{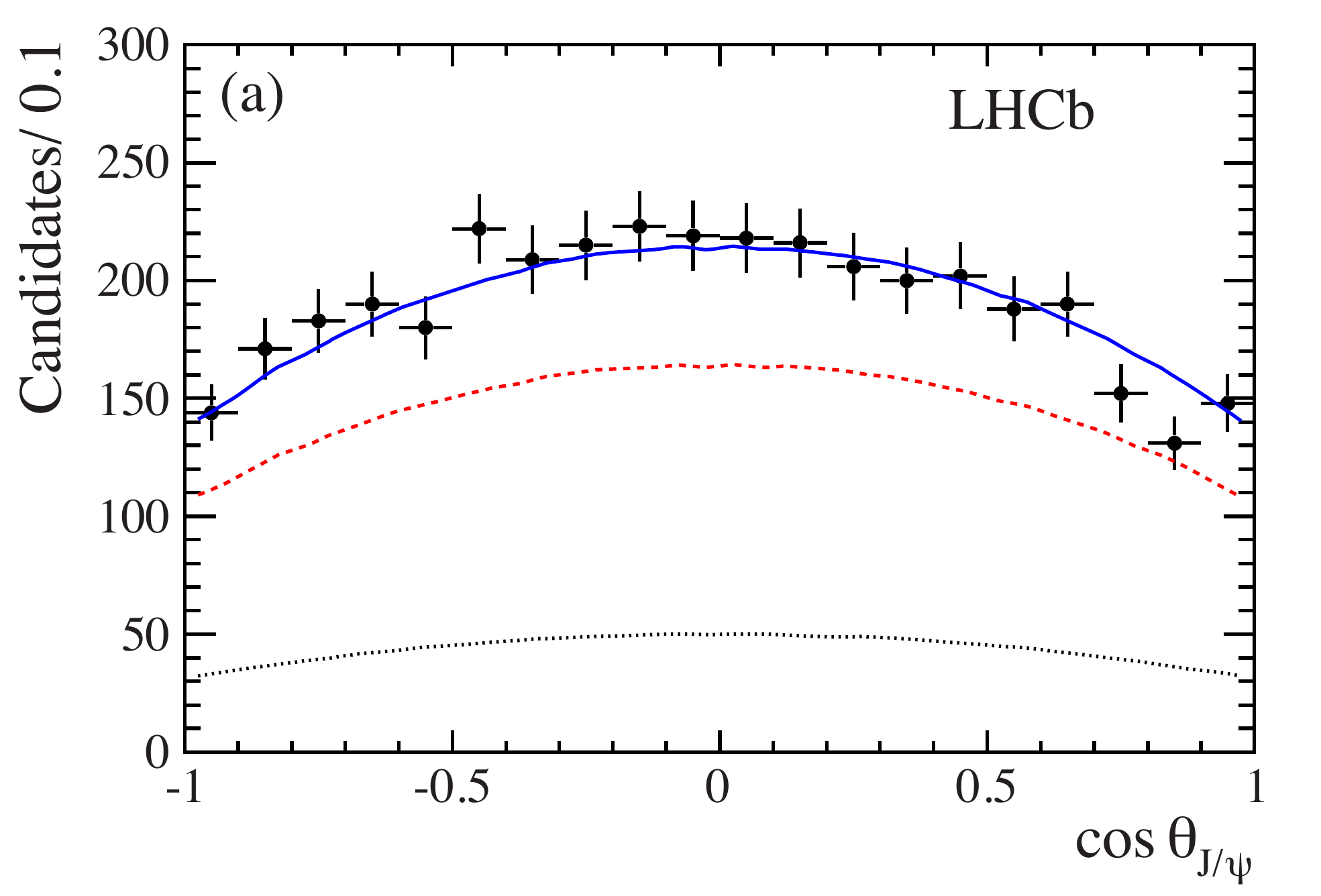}%
    \includegraphics[width=0.48\textwidth]{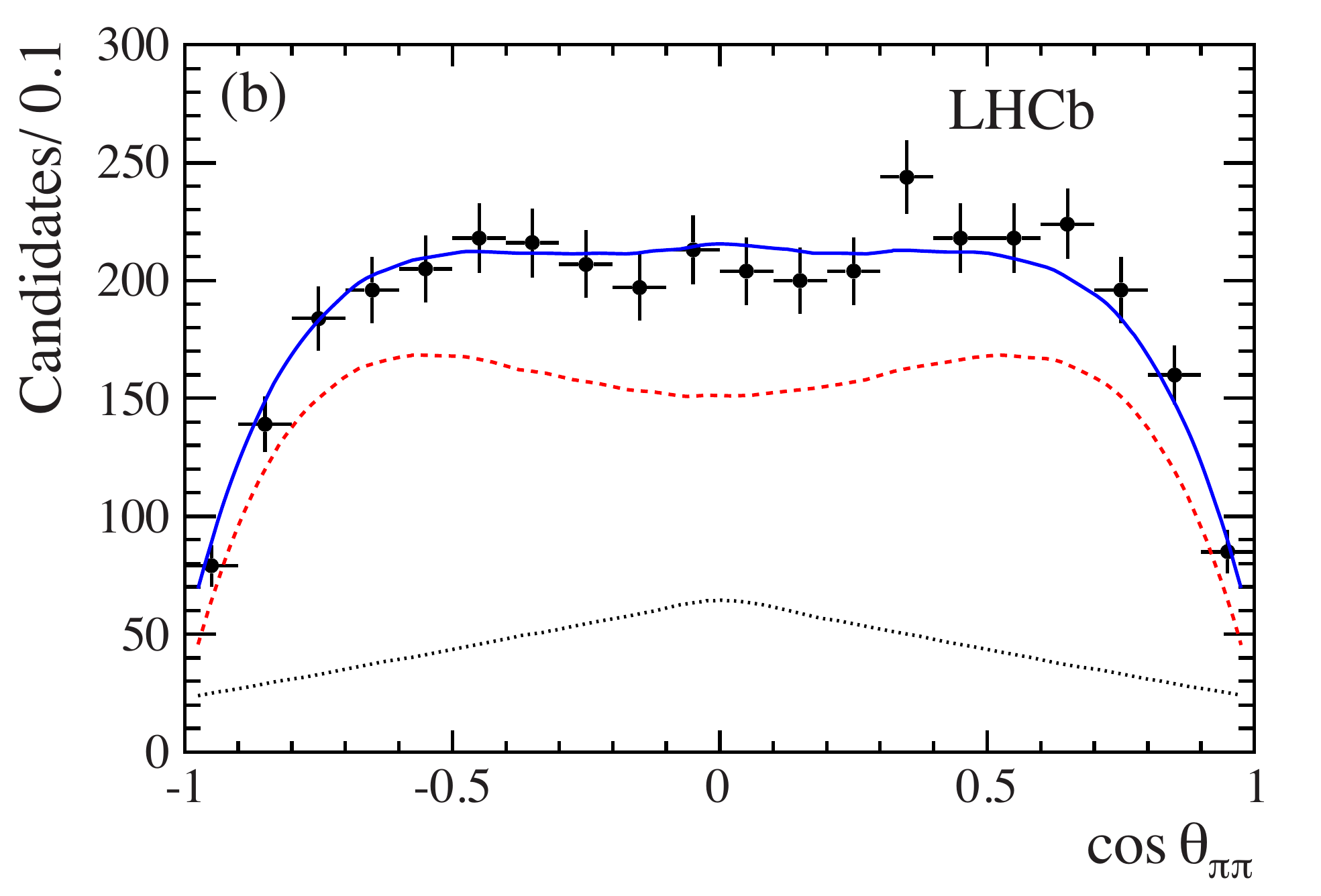}
\end{center}
\vspace{-6mm}
\caption{Helicity angle distributions of (a) $\cos \theta_{\jpsi}$ ($\chi^2$/ndf =15/20) and (b) $\cos\theta_{\pi\pi}$ ($\chi^2$/ndf =14/20) in the $\rho(770)$ mass region defined within one full width of the $\rho(770)$ mass. The points with error bars are data, the signal fit to the best model is shown with a (red) dashed line, the background with a (black) dotted line, and the (blue) solid line represents the total.}\label{plot:hel}
\end{figure}
\begin{figure}[htb]
\begin{center}
    \includegraphics[width=0.48\textwidth]{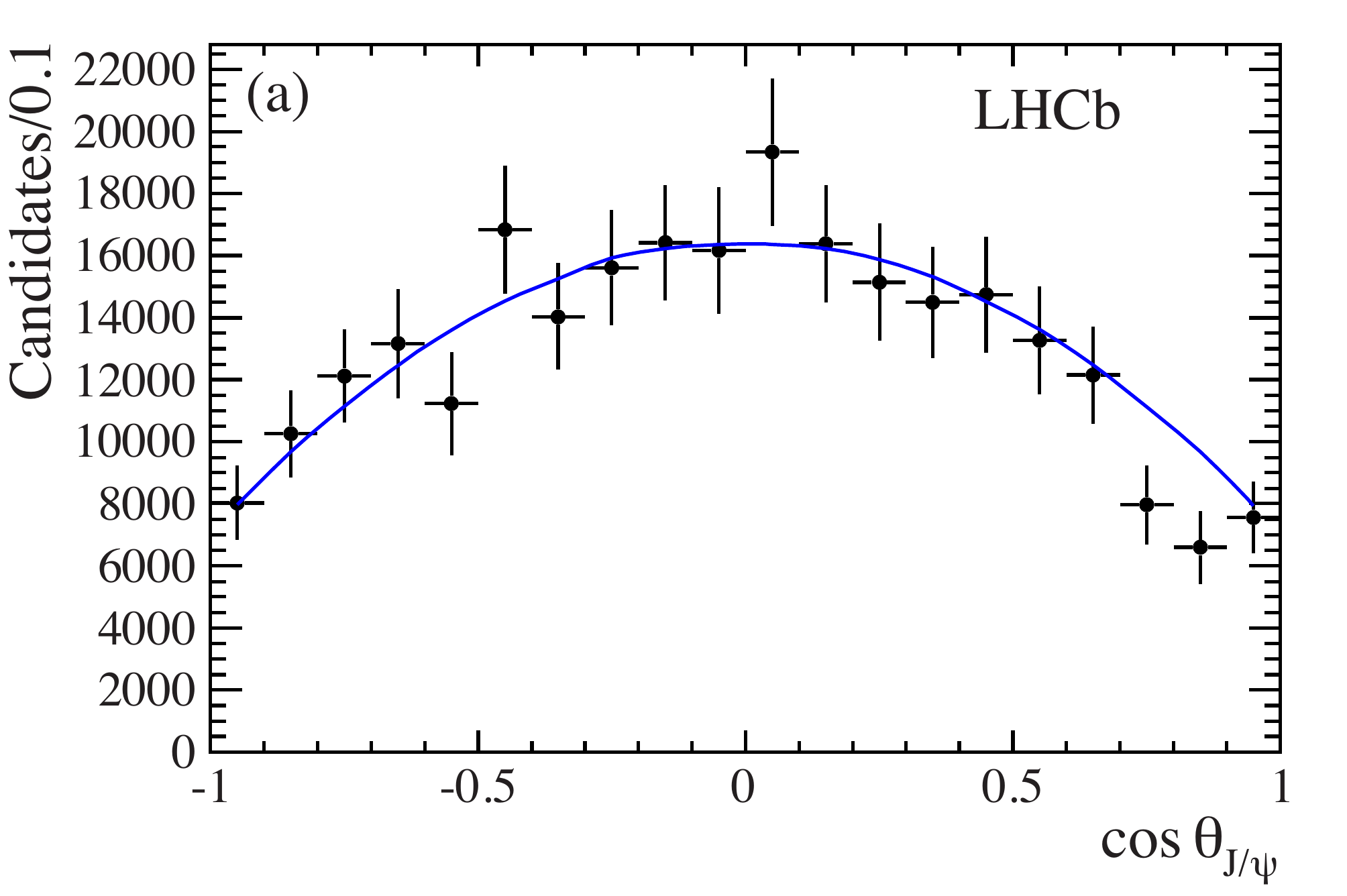}%
    \includegraphics[width=0.48\textwidth]{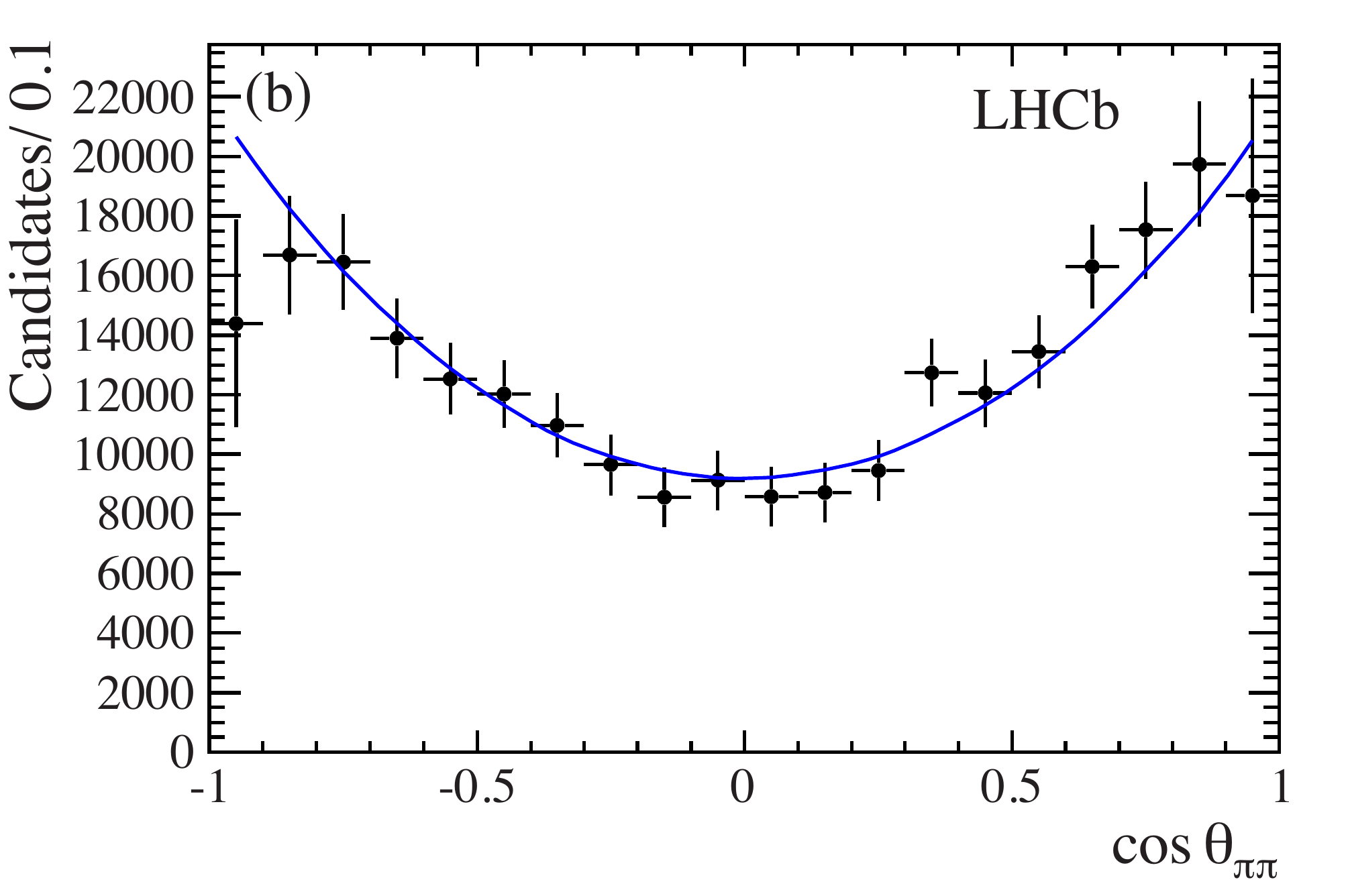}
\end{center}
\vspace{-6mm}
\caption{Background subtracted and efficiency corrected helicity distributions of (a) $\cos \theta_{\jpsi}$ ($\chi^2$/ndf =20/20) and (b) $\cos\theta_{\pi\pi}$ ($\chi^2$/ndf =13/20) in the $\rho(770)$ mass region defined within one full width of the $\rho(770)$ mass.  The points with error bars  are data and the solid blue lines show the fit to the best model.}\label{helii1_4}
\end{figure}

\section{Branching fractions}
Branching fractions are measured by normalizing to the well measured decay mode $B^-\to \jpsi K^-$, which has  two muons in the final state and has the same triggers as the $\Bdb\rightarrow \jpsi \pi^+\pi^-$ decays. Assuming equal production of charged and neutral $B$ mesons at the LHC due to isospin symmetry, the branching fraction is calculated as
\begin{equation}
{\cal B}(\Bzb \to \jpsi \pi^+\pi^-) = \frac{N_{\Bzb} / \epsilon_{\Bzb}}{N_{B^-} / \epsilon_{B^-}}\times {\cal B}(B^- \to \jpsi K^-),
\end{equation}
where $N$ and $\epsilon$ denote the yield and total efficiency of the decay of interest.
The branching fraction ${\cal B}(B^-\to \jpsi K^-)=(10.18\pm0.42)\times10^{-4}$ is determined from an average of recent Belle \cite{Abe:2002rc} and \babar \cite{Aubert:2004rz} measurements that are corrected with respect to the reported values, which assume equal production of charged and neutral $B$ mesons at the $\Upsilon(4S)$, using the measured value of $\frac{\Gamma(B^+B^-)}{\Gamma(\Bz \Bzb)}=1.055\pm0.025$ \cite{Amhis:2012bh}.

Signal efficiencies are derived from simulations including trigger, reconstruction, and event selection components.
Since the efficiency to detect the $\jpsi \pi^+\pi^-$ final state is not uniform across the Dalitz plane, the efficiency is averaged according to the Dalitz model, where the best fit model is used. The $\KS$ veto efficiency is also taken into account. Small corrections are applied to account for differences between the simulation and the data.  We measure the kaon and pion identification efficiencies with respect to the  simulation using  $D^{*+}\to\pi^+\Dz(\to K^-\pi^+)$ events selected from data.  The efficiencies are measured in bins of $\pt$ and $\eta$ and the averages are weighted using the signal event distributions in the data. Furthermore, to ensure that the $p$ and $\pt$ distributions of the generated $B$
mesons are correct we weight the $B^-$ and $\Bzb$ simulation samples using $B^-\to \jpsi K^-$ and $\Bzb\to \jpsi \Kstarzb$ data, respectively. Finally, the simulation samples are weighted with the charged tracking efficiency ratio between data and simulation in bins of $p$ and $\pt$ of the track. The average of the weights is the correction factor. The total correction factors are below 1.04 and largely cancel between the signal and normalization channels.  Multiplying the simulation efficiencies and correction factors gives the total efficiency ($1.163\pm0.003\pm0.017$)\% for $\Bzb\to \jpsi \pi^+\pi^-$ and ($3.092\pm0.012\pm0.038$)\% for $\Bm \to \jpsi K^-$, where the first uncertainty is statistical and the second is systematic.


Using $N_{B^-}=350{,}727\pm633$ and $N_{\Bzb}=5287\pm112$,  we measure
\begin{equation}
{\cal B}(\Bzb \to \jpsi \pi^+\pi^-) = (3.97\pm0.09\pm0.11\pm0.16)\times 10^{-5}, \nonumber
\end{equation}
where the first uncertainty is statistical, the second is systematic and the third is due to the uncertainty of ${\cal B}(B^- \to \jpsi K^-)$. The systematic uncertainties are discussed in Section \ref{sec:syst}. Our measured value is consistent with and more precise than the previous \babar measurement of $(4.6\pm0.7\pm0.6)\times10^{-5}$ \cite{Aubert:2002vb}.

Table~\ref{tab:br} shows the branching fractions of resonant modes calculated by multiplying the fit fraction and the total branching fraction of $\Bzb \to \jpsi \pi^+\pi^-$. Since the $f_0(980)$ contribution has a significance of less than 3$\sigma$ we quote also an upper limit of
${\cal{B}}\left(\Bzb\to \jpsi f_0(980)\right)\times{\cal{B}}\left(f_0(980)\to \pi^+\pi^-\right) < 1.1\times 10^{-6}$
at 90\% confidence level (CL); this is the first such limit. 
 The limit is calculated assuming a Gaussian distribution as the central value plus 1.28 times the addition in quadrature of the statistical and systematic uncertainties.
This branching ratio is predicted to be in the range $(1-3)\times 10^{-6}$ if the $f_0(980)$ resonance  is formed of tetra-quarks,  but can be much smaller if the $f_0(980)$ is a standard quark anti-quark resonance \cite{Fleischer:2011au}. Our limit is at the lower boundary of the tetra-quark prediction, and is consistent with a quark anti-quark resonance with a small mixing angle. In Section~\ref{sec:mixing angle}, we show that the mixing angle, describing the admixture of $s\bar{s}$ and light quarks, is less than 31$^{\circ}$ at 90\% CL.

The other branching fractions are consistent with and more precise than the previous measurements from \babar \cite{Aubert:2002vb,*Aubert:2007xw}. Using ${\cal B}(\omega\to\pi^+\pi^-)=(1.53^{+0.11}_{-0.13})\%$ \cite{Beringer:2012}, we measure
\begin{equation}
\frac{{\cal B}(\Bzb \to \jpsi \omega)}{{\cal B}(\Bzb \to \jpsi \rho^0)}=0.61^{+0.24+0.31}_{-0.14-0.16}, \nonumber
\end{equation}
and
\begin{equation}
{\cal B}(\Bzb \to \jpsi \omega)=(1.5_{-0.3-0.4}^{+0.6+0.7})\times10^{-5}. \nonumber
\end{equation}
This is consistent with the LHCb measurement $\frac{{\cal B}(\Bzb \to \jpsi \omega)}{{\cal B}(\Bzb \to \jpsi \rho^0)}=0.89\pm0.19^{+0.07}_{-0.13}$, using the $\omega\to \pi^+\pi^-\pi^0$ mode \cite{:2012cw}.
\begin{table}[b]
\centering
\caption{Branching fractions for each channel. The upper limit at 90\% CL is also quoted for the $f_0(980)$ resonance which has a significance smaller than 3$\sigma$. The first uncertainty is statistical and the second the total systematic.}
\def\arraystretch{1.2}
\begin{tabular}{lrc}
\hline
Channel & ${\cal B}(\Bzb \to \jpsi R, R \to \pi^+\pi^-)$ & Upper limit of $\cal {B}$ \\
&& (at 90\% CL)\\
\hline
$\rho(770)$& $(2.49_{-0.13-0.23}^{+0.20+0.16})\times10^{-5}$&-\\
$\omega(782)$& $(2.3_{-0.5-0.6}^{+0.9+1.1})\times10^{-7}$& -\\
$f_0(980)$& $(6.1^{+3.1+1.7}_{-2.0-1.4})\times10^{-7}$&$<1.1\times10^{-6}$\\
$f_2(1270)$& $(3.5\pm0.4\pm0.4)\times10^{-6}$&-\\
$\rho(1450)$&$(2.1^{+1.0+2.2}_{-0.6-0.4})\times10^{-6}$&-\\
$f_0(500)$&$(6.4\pm0.8^{+2.4}_{-0.8})\times10^{-6}$&-\\
\hline
\end{tabular}\label{tab:br}
\def\arraystretch{1.0}
\end{table}

\section{Systematic uncertainties}\label{sec:syst}
\begin{table}[b]
\centering
\caption{Relative systematic uncertainties on branching fractions~(\%).}
\vspace{0.2cm}
\begin{tabular}{lc}\hline
Source& Uncertainty (\%)\\\hline

\hline
Tracking efficiency & 1.0 \\
Material and physical effects &2.0 \\
Particle identification efficiency & 1.0 \\
$\Bzb$ $p$ and $\pt$ distributions & 0.5 \\
$\Bm$ $p$ and $\pt$ distributions & 0.5 \\
Dalitz modeling & 0.6\\
Background modeling & 0.5\\
\hline
Sum of above sources &2.7\\
\hline
$\mathcal{B}(\Bm\to \jpsi\Km)$&4.1\\
\hline
Total&4.9\\\hline
\end{tabular}
\label{tab:sys_br}
\end{table}

The contributions to the systematic uncertainties on the branching
fractions are listed in Table~\ref{tab:sys_br}. Since the branching fractions are measured with respect to the $\Bm\to \jpsi\Km$ mode, which has a different number of charged tracks than the decays of interest, a 1\% systematic uncertainty is assigned due to differences in the tracking performance between data and simulation. Another 2\% uncertainty is assigned  because of the difference between two pions and one kaon in the final states, due to decay in flight, multiple scattering, and hadronic interactions. Small uncertainties are introduced if the simulation does not have the correct $\B$ meson kinematic distributions. We are relatively insensitive  to any differences in the $\B$ meson $p$ and $\pt$ distributions since we are measuring the relative rates. By varying the $p$ and $\pt$ distributions we see at most a change of 0.5\%. There is a 1.0\% systematic uncertainty assigned for the relative particle identification efficiencies (0.5\% per particle). These efficiencies have been corrected from those predicted in the simulation by using the data from $D^{*+}\to \pip D^0(\to \Km\pip)$. A 0.6\% uncertainty is included for the $\jpsi\pi^-\pi^+$ efficiency, estimated by changing the best model to that including all possible resonances. The $\Bzb$ signal yield is changed by 0.5\% when the shape of the
combinatorial background is changed from an exponential to a linear function. The total systematic uncertainty is obtained by adding each source of systematic uncertainty in quadrature as they are uncorrelated. In addition, the largest source is $4.1\%$ due to the uncertainty of $\mathcal{B}(\Bm\to \jpsi\Km)$ which is quoted separately.

\begin{table}[t]
\centering
\caption{Absolute systematic uncertainties on the results of the Dalitz analysis.}
\def\arraystretch{1.2}
\begin{tabular}{lccccc}
\hline
Item& Acceptance& Background& Fit model & Resonance parameters& Total\\\hline
\multicolumn{6}{c}{Fit fractions (\%)}\\\hline
$\rho(770)$& $\pm0.9$   &$^{+2.0}_{-3.1}$          &$^{+0.6}_{-3.2}$    &$\pm1.6$&$^{+2.8}_{-4.8}$\\
$\omega(782)$&$\pm0.08$&$^{+0.23}_{-0.06}$      &$^{+0.11}_{-0.10}$&$^{+0.028}_{-0.014}$&$_{-0.14}^{+0.27}$\\
$f_0(980)$& $\pm0.03$   &$_{-0.17}^{+0.24}$       &$^{+0.21}_{-0.18}$  &$_{-0.24}^{+0.29}$&$^{+0.43}_{-0.35}$\\
$f_2(1270)$& $\pm0.06$  &$_{-0.59}^{+0.45}$       &$^{+0.85}_{-0.76}$  &$\pm0.36$&$\pm1.0$\\
$\rho(1450)$&$\pm0.10$  &$^{+0.5}_{-0.6}$        &$^{+5.6}_{-0.7}$    &$^{+0.4}_{-0.3}$&$^{+5.6}_{-0.9}$\\
$f_0(500)$& $\pm0.4$    &$_{-0.9}^{+1.6}$        &$^{+5.7}_{-1.6}$    &$\pm0.6$&$^{+6.0}_{-2.0}$\\
\hline
\multicolumn{6}{c}{$\lambda=0$ fractions (\%)}\\\hline
$\rho(770)$&$\pm1.0$&$_{-2.0}^{+1.7}$&$^{+4.9}_{-1.5}$&$\pm2.1$&$^{+5.7}_{-3.4}$\\
$\omega(782)$&$\pm1.5$&$_{-1.8}^{+3.5}$&$^{+2.8}_{-3.5}$&$_{-1.7}^{+1.2}$&$^{+4.9}_{-4.5}$\\
$f_2(1270)$&$\pm0.3$&$\pm2.4$&$^{+3.7}_{-3.4}$&$\pm1.5$&$\pm4.5$\\
$\rho(1450)$&$\pm0.9$&$_{-8.4}^{+4.8}$&$^{+5.5}_{-5.1}$&$^{+4.2}_{-6.1}$&$^{+8.4}_{-11.6}$\\
\hline
\multicolumn{6}{c}{Ratio of fit fractions (\%)}\\\hline
$\omega(782)/\rho(770)$&{$\pm0.13$}&{$_{-0.11}^{+0.41}$}&{$_{-0.16}^{+0.18}$}&{$_{-0.022}^{+0.034}$}&{$^{+0.47}_{-0.23}$}\\
$f_0(980)/f_0(500)$&$\pm0.3$&$^{+1.5}_{-1.1}$&$^{+1.0}_{-2.1}$&$_{-1.8}^{+2.2}$&$\pm3.0$\\
\hline
\end{tabular}
\def\arraystretch{1.0}
\label{sys:dlz}
\end{table}

The sources of the systematic uncertainties on the results of the Dalitz plot analysis are summarized in Table \ref{sys:dlz}. For the uncertainties due to the acceptance or background modeling, we repeat the data fit 100 times where the parameters of acceptance or background modeling are generated according to the corresponding covariance matrix. We also study the acceptance function by changing the minimum IP $\chi^2$ requirement from 9 to 12.5 on both of the pion candidates. As shown previously~\cite{LHCb:2012ae}, this increases the $\chi^2$ of the fit to the angular distributions by one unit. The acceptance function is then applied to the data with the original minimum IP $\chi^2$ selection of 9, and the likelihood fit is redone and the uncertainties are estimated by comparing the results with the best fit model. The larger of the two variations is taken as uncertainty due to the acceptance.

We study the effect of ignoring the experimental mass resolution in the fit by comparing fits between different pseudo-experiments with and without the resolution included. As the widths of the resonances we consider are much larger than the mass resolution, we find that the effects are negligible except for the $\omega(782)$  resonance whose fit fraction is underestimated by ($0.09\pm0.08$)\%. Thus, we apply a $0.09\%$ correction to the $\omega(782)$ fraction and assign an additional $\pm0.08\%$ in the acceptance systematic uncertainty. The results shown in the previous sections already include this correction.

In the default fit, the signal fraction $f_{\rm sig}=0.621\pm 0.009$, defined in Eq. \ref{eq:pdf} is fixed; we vary its value within its error to estimate the systematic uncertainty. The change is added in quadrature with the background modeling uncertainties.

The uncertainties due to the fit model include adding each resonance that is listed in Table \ref{tab:resparam} but not used in the best model, changing the default values of $L_B$ in P- and D-wave cases, varying the hadron scale $r$ parameters for  the $B$ meson and $R$ resonance to $3.0\gev^{-1}$ for both, replacing the $f_0(500)$ model by a Zhou and Bugg function \cite{Bugg:2003kj,Ablikim:2004qna} and using the alternate Gounaris and Sakurai model \cite{GS} for $\rho$ resonances. Then the largest variations among those changes are assigned as the systematic uncertainties for modeling (see Table~\ref{sys:dlz}).

Finally, we repeat the data fit by varying the mass and width of resonances (see Table~\ref{tab:resparam}) within their errors one at a time, and add the changes in quadrature.

\section{Further results and implications}
\subsection{Resonant structure}
The largest intermediate state in $\Bzb \to \jpsi \pi^+\pi^-$ decays is the $\jpsi\rho(770)$ mode. Beside the $\rho(770)$, significant $f_2(1270)$ and $f_0(500)$ contributions are also seen. The smaller $\omega(782)$ and $\rho(1450)$ resonances have 3.1$\sigma$ and 3.2$\sigma$ significances respectively, including systematic uncertainties. The systematic uncertainties reduce the significance of the $f_0(980)$ to below $3\sigma$. Replacing the $f_0(500)$ by a non-resonant component increases $-\ln{\cal L}$ by 117, and worsens the $\chi^2$ by 192 with the same ndf resulting in a fit confidence level of $1.8\times10^{-7}$. Thus the $f_0(500)$ state is firmly established in $\Bzb \to \jpsi \pi^+\pi^-$ decays.

As discussed in the introduction, a region with only S- and P-waves is preferred for measuring $\sin2\beta^{\rm eff}$. The best fit model demonstrates that the mass region within $\pm149$\mev (one full width) of the $\rho(770)$ mass contains only $(0.72\pm0.09)\%$ D-wave contribution, thus this region can be used for  a clean \CP measurement. The S-wave in this region is (11.9$\pm$1.7)\%, where the fraction is the sum of individual fit fractions and the interference.

\subsection{Mixing angle between \boldmath{$f_0(980)$} and $f_0(500)$}
\label{sec:mixing angle}
The scalar nonet is quite an enigma. The mysteries are summarized in Ref. \cite{Schechter:2012zc}, and in the ``Note on scalar mesons" in the PDG \cite{Beringer:2012}.  Let us contrast the masses of the lightest vector mesons with those of the scalars, listed in Table~\ref{tab:nonet}.
\begin{table}[t]
\centering
\caption{Masses of light vector and scalar resonances. All values are taken from \cite{Beringer:2012}, except for the $f_0(500)$ \cite{Muramatsu:2002jp}.}
\begin{tabular}{ccccc}
\hline
Isospin & Vector particle& Vector mass (\mev)& Scalar particle & Scalar mass (\mev)\\
\hline
0 & $\omega$ &783 & $f_0(500)$ & 513\\
1& $\rho$ &776 & $a_0$ & 980  \\
1/2 & $K^*$ & 980 &$\kappa$ &800 \\
0 & $\phi$ & 1020 & $f_0$ & 980 \\
\hline
\end{tabular}\label{tab:nonet}
\end{table}
For the vector particles, the $\omega$ and $\rho$ masses are nearly degenerate and the masses increase as the $s$-quark content increases. For the scalar particles, however, the mass dependence differs in several ways which requires an explanation. Some authors introduce the concept of $q\bar{q}q\bar{q}$ states or superpositions of the four-quark state with the $q\bar{q}$ state. In either case,  the $I=0$ $f_0(500)$ and the $f_0(980)$ are thought to be mixtures of the underlying states whose mixing angle has been estimated previously (see Ref.~\cite{Fleischer:2011au} and references contained therein).

The mixing is parameterized by a 2$\times$2 rotation matrix characterized by the angle $\varphi_m$, giving in our case
\begin{eqnarray}
  \label{eq:fmix}
 \ket{f_0(980)}&=&\;\;\;\cos\varphi_m\ket{s\bar{s}}+\sin\varphi_m\ket{n\bar{n}}\nonumber\\
  \ket{f_0(500)}&=&-\sin\varphi_m\ket{s\bar{s}}+\cos\varphi_m\ket{n\bar{n}},\nonumber\\
  {\rm where~} \ket{n\bar{n}}&\equiv&\frac{1}{\sqrt{2}}\left(\ket{u\bar{u}}+\ket{d\bar{d}}\right).
\end{eqnarray}

In this case only the $\ket{d\bar{d}}$ part of the $\ket{n\bar{n}}$ wave function contributes (see Fig.~\ref{feyn3}).  Thus we have
\begin{equation}
\tan^2\varphi_m
=\frac{{\cal{B}}\left(\Bzb\to\jpsi f_0(980)\right)}{{\cal{B}}\left(\Bzb\to\jpsi f_0(500)\right)}\frac{\Phi(500)}{\Phi(980)},
\end{equation}
where the $\Phi$ terms denote the phase space factors. The phase space in this pseudoscalar to vector-pseudoscalar decay is proportional to the cube of the $f_0$ three-momentum. Taking the average of the momentum dependent phase space over the resonant line shapes results in the ratio of phase space factors $\frac{\Phi(500)}{\Phi(980)}$ being equal to 1.25.

Using the data shown in Table~\ref{tab:br}  we determine the ratio of branching fractions for both resonances resulting in the $\pi^+\pi^-$ final state as
\begin{equation}
\frac{{\cal{B}}\left(\Bzb\to\jpsi f_0(980)\right)\times{\cal{B}}\left(f_0(980)\to\pi^+\pi^-\right)}{{\cal{B}}\left(\Bzb\to\jpsi f_0(500)\right)\times{\cal{B}}\left(f_0(500)\to\pi^+\pi^-\right)}=(9.5^{+6.7}_{-3.4}\pm3.0)\%. \nonumber
\end{equation}
This value must be corrected for the individual branching fractions of the $f_0$ resonances into the $\pi^+\pi^-$ final state.

\babar has measured $\frac{{\cal{B}}\left(f_0(980)\to K^+K^-\right)}{{\cal{B}}\left(f_0(980)\to \pi^+\pi^-\right)}=0.69\pm0.32$
using $B\to KKK$ and $B\to K\pi\pi$ decays \cite{Aubert:2006nu}. BES obtained relative branching ratios using $\psi(2S)\to\gamma \chi_{c0}$ decays where the $\chi_{c0}\to f_0(980)f_0(980)$, and either both $f_0(980)$ candidates decay into $\pi^+\pi^-$ or one into $\pi^+\pi^-$ and the other into $K^+K^-$ pairs \cite{Ablikim:2004cg,*Ablikim:2005kp}. From their results we obtain $\frac{{\cal{B}}\left(f_0(980)\to K^+K^-\right)}{{\cal{B}}\left(f_0(980)\to \pi^+\pi^-\right)}=0.25^{+0.17}_{-0.11}$\cite{Ecklund:2009aa}. Averaging the two measurements gives
\begin{equation}
\frac{{\cal{B}}\left(f_0(980)\to K^+ K^-\right)}{{\cal{B}}\left(f_0(980)\to\pi^+\pi^-\right)}=0.35_{-0.14}^{+0.15}\,.
\end{equation}

Assuming that the $\pi\pi$ and $KK$ decays are dominant we obtain
\begin{equation}
{\cal{B}}\left(f_0(980)\to\pi^+\pi^-\right)
=\left(46\pm6\right)\%, 
\end{equation}
where we have assumed that the only other decays are to $\pi^0\pi^0$, half of the $\pi^+\pi^-$ rate, and to neutral kaons, taken equal to charged kaons. We use ${\cal{B}}\left(f_0(500)\to\pi^+\pi^-\right)
=\frac{2}{3}$, which results from isospin Clebsch-Gordon coefficients, and assuming that the only decays are into two pions. Since we have only an upper limit on the $\jpsi f_0(980)$ final state, we will only find an upper limit on the mixing angle, so if any other decay modes of the $f_0(500)$ ($f_0(980)$) exist, they would make the limit more (less) stringent. Our limit then is
\begin{equation}
\tan^2\varphi_m=\frac{{\cal{B}}\left(\Bzb\to\jpsi f_0(980)\right)}{{\cal{B}}\left(\Bzb\to\jpsi f_0(500)\right)}\frac{\Phi(500)}{\Phi(980)} <0.35~{\rm at~90\%~confidence~level}, \nonumber
\end{equation}
which translates into a limit
\begin{equation}
|\varphi_m| < 31^{\circ}~{\rm at~90\%~confidence~level}. \nonumber
\end{equation}

Various mixing angle measurements have been derived in the literature and summarized in Ref.~\cite{Fleischer:2011au}. There are a wide range of values including:  (a) using $D_s^+\to\pi^+\pi^+\pi^-$ transitions which give a range $35^{\circ}<|\varphi_m|<55^{\circ}$, (b)
using radiative decays where two solutions were found either $\varphi_m=4^{\circ}\pm3^{\circ}$ or $136^{\circ}\pm 6^{\circ}$, (c)
using resonance decays from both $\phi\to\gamma\pi^0\pi^0$ and $\jpsi\to\omega\pi\pi$ where a value of $\varphi_m\simeq 20^{\circ}$ was found, (d)
using the $D^{\pm}$ and $D_s^{\pm}$ decays into $f_0(980)\pi^{\pm}$ and $f_0(980)K^{\pm}$ where values of $\varphi_m=31^{\circ}\pm5^{\circ}$ or $42^{\circ}\pm 7^{\circ}$ were found.

\section{Conclusions}
We have studied the resonance structure of $\Bzb\rightarrow \jpsi \pi^+\pi^-$ using a modified Dalitz plot analysis where we also include the decay angle of the $\jpsi$ meson. The decay distributions are formed from a series of final states described by individual $\pi^+\pi^-$ interfering decay amplitudes. The largest component is the $\rho(770)$ resonance. The data are best described by adding the $f_2(1270)$, $f_0(500)$, $\omega(782)$, $\rho(1450)$ and $f_0(980)$ resonances, where the $f_0(980)$ resonance contributes less than $3\sigma$ significance.  The results are listed in Table \ref{ff3}. %

We set an upper limit ${\cal{B}}\left(\Bzb\to \jpsi f_0(980)\right)\times{\cal{B}}\left(f_0(980)\to \pi^+\pi^-\right) < 1.1 \times 10^{-6}$ at 90\% confidence level that favors somewhat a quark anti-quark interpretation of the $f_0(980)$ resonance. We also have firmly established the existence of the $\jpsi f_0(500)$ intermediate resonant state in \Bzb decays, and limit the absolute value of the mixing angle between the two lightest scalar states to be less than $31^{\circ}$ at 90\% confidence level.

Our six-resonance best fit shows that the mass region within one full width of the $\rho(770)$ contains mostly P-wave, $(11.9\pm1.7)$\% S-wave, and only $(0.72\pm0.09)$\% D-wave. Thus this region can be used to perform \CP violation measurements, as the S- and P-wave components can be treated in the same manner as in the analysis of $\Bsb\to\jpsi\phi$ \cite{LHCb-CONF-2012-002,LHCb:2011aa,Aaltonen:2012ie,*Abazov:2011ry,*:2012fu}. The measured value of the asymmetry can be compared to that found in other modes such as $\Bzb\to \jpsi \Kzb$ in order to ascertain the possible effects due to penguin amplitudes.


The measured  branching ratio is
\begin{equation}
{\cal B}(\Bzb \to \jpsi \pi^+\pi^-) = (3.97\pm0.09\pm0.11\pm0.16)\times 10^{-5}, \nonumber
\end{equation}
where the first uncertainty is statistical, the second is systematic and the third is due to the uncertainty of ${\cal B}(B^- \to \jpsi K^-)$.
The largest contribution is the $\jpsi\rho(770)$  mode with a branching fraction of $(2.49_{-0.13-0.23}^{+0.20+0.16})\times10^{-5}$.
\section*{Acknowledgements}

\noindent We express our gratitude to our colleagues in the CERN
accelerator departments for the excellent performance of the LHC. We
thank the technical and administrative staff at the LHCb
institutes. We acknowledge support from CERN and from the national
agencies: CAPES, CNPq, FAPERJ and FINEP (Brazil); NSFC (China);
CNRS/IN2P3 and Region Auvergne (France); BMBF, DFG, HGF and MPG
(Germany); SFI (Ireland); INFN (Italy); FOM and NWO (The Netherlands);
SCSR (Poland); ANCS/IFA (Romania); MinES, Rosatom, RFBR and NRC
``Kurchatov Institute'' (Russia); MinECo, XuntaGal and GENCAT (Spain);
SNSF and SER (Switzerland); NAS Ukraine (Ukraine); STFC (United
Kingdom); NSF (USA). We also acknowledge the support received from the
ERC under FP7. The Tier1 computing centres are supported by IN2P3
(France), KIT and BMBF (Germany), INFN (Italy), NWO and SURF (The
Netherlands), PIC (Spain), GridPP (United Kingdom). We are thankful
for the computing resources put at our disposal by Yandex LLC
(Russia), as well as to the communities behind the multiple open
source software packages that we depend on.

\ifx\mcitethebibliography\mciteundefinedmacro
\PackageError{LHCb.bst}{mciteplus.sty has not been loaded}
{This bibstyle requires the use of the mciteplus package.}\fi
\providecommand{\href}[2]{#2}

\end{document}